\documentclass{scrartcl}

\KOMAoptions{%
    abstract=true,
    fontsize=10pt,
    footinclude=false,
    headinclude=false,
    headsepline=false,
    mpinclude=false,
    numbers=noenddot,
    paper=letter,
    parskip=half-,
    twocolumn=true,
    twoside=false,
}

\usepackage{checkend}
\usepackage{etex}
\usepackage{etoolbox}
\usepackage{hyperref}
\usepackage{xcolor}
\usepackage{xifthen}
\usepackage{xspace}

\makeatletter
\providecommand{\institute}[1]{
    \apptocmd{\@author}{\end{tabular}\par\begin{tabular}{c}#1}{}{}
}
\makeatother

\definecolor{href:citecolor}{rgb}{.002,.002,.56}
\definecolor{href:linkcolor}{rgb}{.002,.002,.56}

\AtEndPreamble{%
  \hypersetup{%
    breaklinks=true,%
    citecolor=href:citecolor,%
    colorlinks=true,%
    linkcolor=href:linkcolor,%
    plainpages=false,%
    urlcolor=href:linkcolor,%
    pdfinfo = {%
      Title    = {\@title},%
      Subject  = {},%
      Author   = {},%
      URI      = {},%
      Keywords = {},%
    }%
  }
}

\newboolean{authoryear}
\setboolean{authoryear}{true}

\ifthenelse{\boolean{authoryear}}{%
    \usepackage[%
        backend=biber,
        dashed=false,
        sorting=anyvt,
        style=ext-authoryear, 
    ]{biblatex}
    \DeclareOuterCiteDelims{parencite}{\bibopenbracket}{\bibclosebracket}
}{%
    \usepackage[%
        backend=biber,
        sorting=none,
        style=numeric-comp,
    ]{biblatex}
}

\usepackage{blindtext}
\usepackage{pbalance}
\usepackage{relsize}
\usepackage{url}

\usepackage[extendedchars,space]{grffile}
\usepackage[export]{adjustbox}
\usepackage{subfig}

\usepackage{multirow}
\newcommand{\mrow}[3]{\multirow{#1}{#2}{#3}}

\usepackage{units}
\usepackage{xspace}

\newcommand{\kbps}[1]{{\unit[#1]{kbps}}\xspace}

\newcommand{\HZ}  [1]{{\unit[#1]{Hz}}\xspace}
\newcommand{\KHZ} [1]{{\unit[#1]{kHz}}\xspace}
\newcommand{\MIN} [1]{{\unit[#1]{min}}\xspace}
\newcommand{\DB}  [1]{{\unit[#1]{dB}}\xspace}

\newcommand{\CMD}[1]{{\ttfamily #1}}

\newcommand{\eg}{{e.g.,}\xspace}
\newcommand{\ie}{{i.e.}\xspace}
\newcommand{\cf}{{cf.}\xspace}

\newcommand{\textplus}{%
    \raisebox{\dimexpr(\fontcharht\font`X-\height+\depth)/2\relax}{+}\xspace
}

\usepackage{booktabs}
\usepackage{array}
\usepackage{nicefrac}
\usepackage{nicematrix}
\usepackage{tabularx}

\newcolumntype{L}[1]{>{\hsize=#1\hsize\raggedright\arraybackslash}X}%
\newcolumntype{R}[1]{>{\hsize=#1\hsize\raggedleft\arraybackslash}X}%
\newcolumntype{C}[1]{>{\hsize=#1\hsize\centering\arraybackslash}X}%

\usepackage{rotating}

\AtEndPreamble{%
    \usepackage[nameinlink,noabbrev]{cleveref}
    \setcounter{topnumber}{5}
    \setcounter{bottomnumber}{5}
    \setcounter{totalnumber}{5}
}

\newcommand{\MMSEST}{\ensuremath{\text{MMS}_{\text{est}}}\xspace}
\newcommand{\TNMR}{{\ttfamily totalNMR}\xspace}

\title{%
    Evaluation of Audio Compression Codecs
}
\author{%
    Thien T. Duong
    \and
    Jan P. Springer
}
\institute{%
    Computer Science, UA Little Rock
}
\date{}

\begin{document}

\maketitle

\abstract{%
    Perceptual quality of audio is the combination of aural accuracy and listener-perceived sound fidelity. It defines how humans respond to the accuracy, intelligibility, and fidelity of aural media. Today this fidelity is also heavily influenced by the use of audio-compression codecs used for storing aural media in digital form. We argue that, when choosing an audio-compression codec, users should not only look at compression efficiency but also consider the sonic perceptual quality properties of available audio-compression codecs.
    \par
    
    We evaluate several audio-compression codecs, traditional as well as AI-based, in terms of compression performance and their sonic perceptual quality via codec-performance measurements, various visualization techniques, and a family of PEAQ scores. We demonstrate how perceptual quality is affected by digital audio-compression techniques, providing insights for a wide range of users in the process of choosing a digital audio-compression scheme.
    \par
}
{\sffamily\bfseries Keywords}: Audio Codecs, Audio Compression, Information Visualization, Quantitative and Qualitative Audio Evaluation, PEAQ

\section{Introduction}
\label{sec:introduction}

Perceptual quality of audio is the combination of aural accuracy and listener-perceived sound fidelity \autocite{Ellis:PRA:1992}. How listeners experience digital audio depends on {\em sound fidelity}, \ie the accuracy of a digital signal representing the reproduction of sound. {\em High-fidelity} or hi-fi digital audio ideally should have low levels of noise and distortion in its signal to provide a high perceptual reproduction quality.
\par

Our goal is to assess widely used audio-compression codecs such as \textcite{FLACWebsite} or AAC and MP3 \autocite{Brandenburg:MP3AACE:1999} as well as AI-based audio codecs such as RVQGAN \autocite{Kumar:HFACIRVQGAN:2023} with respect to compression efficiency as well as how the compression techniques in these audio codecs affect the perceptual quality of the signal reconstruction. We use visual analytics derived techniques to present the perceptual impact of compression techniques via spectrograms, sound-field diagrams, and objective perceptual scoring based on {\em Perceptual Evaluation of Audio Quality} (PEAQ) \autocite{Thiede:PEAQITUSOMPAQ:2000} and its variant implementations \autocite{Kabal:EIITUPEAQ:2002}.
\par

While subjective listening tests are considered the gold standard for audio-quality evaluation, our test method is based on {\em objective} quality scoring using PEAQ. We believe this will allow for more straightforward replication by non-domain experts as well as allowing additional results to be added into community maintained repositories.
\par

We argue that in choosing an audio-compression codec, users should not only look at the compression efficiency but also consider the sonic perceptual quality of an audio-compression codec in reproducing audio signals.
\par

\section{Related Work}
\label{sec:related+work}

Audio codecs use general compression techniques to reduce the size of digital-audio content. Traditional audio compression techniques include Huffman coding \autocite{Huffman:MCMRC:1952,Bosi:IDACS:2003}, Linear Predictive Coding (LPC) \autocite{Liebchen:IMALC:2004}, and psychoacoustic-based perceptual encoding \autocite{Burg:DSMCAS:2016}.
\par

Audio codecs are categorized into {\em lossy} and {\em lossless} depending on the employed compression scheme. Lossy audio codecs, such as MP3 and AAC \autocite{Brandenburg:MP3AACE:1999} or Vorbis \autocite{Moffitt:OVOFA:2001}, compress digital audio by removing {\em unimportant} data, \ie parts of the digital signal that can be difficult to perceive by a listener because it is not within the human hearing threshold. Lossless audio codecs such as \textcite{FLACWebsite} encode digital audio by only removing redundancies and compressing repeated patterns in the audio signal while maintaining all of the original signal's data.
\par

Assessing digital audio codecs requires comparing compression efficiency as well as evaluating the resulting perceptual quality. Compression metrics include compression ratio, encoding speed, and decoding speed. Evaluating perceptual quality requires more complex assessments. 
\par

Classically, perceptual-quality assessments require trained listeners to perform listening tests \autocite{Olive:HHTL:2011}, which is considered  the gold standard in perceptual-quality evaluation. A subjective listening test such as MUSHRA \autocite{ITU:BS.1534-3:2015} requires trained listeners to be accurate. However, it has been proven that biases exist in MUSHRA \autocite{Zielinski:PBMLT:2007}. Additionally, recent media coverage of blind listening tests has brought attention to instances where listeners were unable to discriminate between sound transmitted through significantly different types of conductive materials \autocite{DIYAudioBlindTest}. This emphasizes our view that subjective listening tests do not necessarily match human intuition regarding audio quality. 
\par

We present our perceptual-quality assessment using various visualization techniques as well as objective scoring to evaluate model-predicted perceptual impact. Tools such as frequency spectra \autocite{AudacityWebsite}, spectrograms \autocite{SpekWebsite}, and sound-field diagrams \autocite{IzotopeWebsite} allow exposing the impact of compression artifacts on the perceptual quality of compressed digital audio. Additionally, {\em Perceptual Evaluation of Audio Quality} (PEAQ) \autocite{Thiede:PEAQITUSOMPAQ:2000} is used to quantify the perceptual quality of audio signals by generating a single {\em Objective Difference Grade} (ODG) score of encoded digital audio.
\par

However, PEAQ is not perfect and there are many attempts to improve on the objective evaluation model. Advanced PEAQ \autocite{Kabal:EIITUPEAQ:2002} has been suggested for modern audio codecs, which tend to exhibit better perceptual encoding algorithms. \Textcite{Dick:ODAQ:2025} show that by using the PEAQ's model output variable (MOV) \TNMR by itself is already a good evaluation metric in combination with advanced PEAQ's ODG score. Additionally, the {\em 2f model} \autocite{Kastner:AEMESQ:2019} in particular is a good example for objectively approximating subjective MUSHRA scores \autocite{Delgado:CWSUP:2020}.
\par

In recent years, experiments with audio compression techniques based on machine-learning and deep-learning techniques have been conducted. There have been several AI-based compression technique proposed, which include end-to-end approaches \autocite{Rim:DNNETE:2021}, lightweight-transformer approaches \autocite{Defossez:HFNAC:2022}, and novel audio codecs such as {\em APCodec} \autocite{Ai:APCodec:2024}. We chose \textcite{Kumar:HFACIRVQGAN:2023}'s {\em Residual Vector Quantized Generative Adversarial Networks} (RVQGAN) model because the authors' claim that their model outperforms competing traditional audio compression algorithms.
\par

We do not intend to replace subjective listener based evaluation with our work. Rather we explore how computational-perceptual models such as PEAQ and its variants in combination with a variety of visualization techniques can facilitate reproducible benchmarking and exploratory evaluation in line with psychoacoustic principles for scalable audio-codec analysis.
\par

\section{Experimental Setup}
\label{sec:setup}

\subsection{Hardware and Software}
\label{sec:setup:hardwaresoftware}

All experiments were conducted on a computer workstation featuring an AMD Ryzen 5 5600X CPU, 32GB RAM at 3200 MHz, and a Samsung 970 EVO NVMe SSD storage device. The reference listening setup included a \textcite{DangerousMusicWebsite} digital-to-analog converter (DAC), a \textcite{SangakuDIYWebsite}, and \textcite{SennheiserHD600Website} open-back headphones.
\par

Encoding and decoding of digital audio used \textcite{FMediaWebsite}, a flexible audio processing tool supporting most existing lossy and lossless codecs. Additionally, fmedia also allows recording of resulting file sizes as well as elapsed time for audio encoding and audio decoding processes. These measurements were used to calculate compression ratio, audio encoding speed, and audio decoding speed.
\par

\subsection{Audio Codecs and Settings}
\label{sec:setup:codec}

Audio-compression ratio in percentage is calculated as shown in \cref{eq:setup:codec:compression+ratio}.
\begin{equation}
    \text{Compression ratio}=\frac{\text{Uncompressed file size}}{\text{Compressed file size}}\cdot100
    \label{eq:setup:codec:compression+ratio}
\end{equation}
The lower the compression ratio the more storage space a digital-audio file will occupy. Encoding speed and decoding speed are expressed in samples per microsecond and calculated as shown in \cref{eq:setup:codec:coding+speed}.
\begin{equation}
    \text{Speed}=\frac{2\cdot\text{Sample size}}{\text{Recorded time}\cdot1,000,000}
    \label{eq:setup:codec:coding+speed}
\end{equation}
Sample size in \cref{eq:setup:codec:coding+speed} is multiplied by two because audio tracks (usually) contain stereo signals in a left and a right channel. In addition, calculated speeds are divided by 1,000,000 in \cref{eq:setup:codec:coding+speed} to convert from samples per second to samples per microseconds.
\par

\begin{table}
    \centering\sffamily\smaller
    \begin{tabularx}{\linewidth}{L{0.7}L{0.7}L{1.1}R{1.5}R{1}}
    \toprule
             & \multirow{2}{*}{Codec} & \multicolumn{3}{c}{Option} \\
                                        \cmidrule(lr){3-5}
             &        & \multicolumn{1}{c}{Name} & \multicolumn{1}{c}{Range} & \multicolumn{1}{c}{Setting} \\
    \midrule
    Lossless & FLAC   & Compression & 0\,--\,8                       & level 6 \\
    \midrule
             & MP3    & CBR         & \kbps{64\,--\,320}             & \kbps{320} \\
             & MP3    & CBR         & \kbps{64\,--\,320}             & \kbps{128} \\
    Lossy    & AAC    & CBR         & \kbps{\hspace{.5em}8\,--\,800} & \kbps{256} \\
             & AAC    & VBR         & 1\,--\,5                       & level 5 \\
             & Vorbis & VBR         & 1\,--\,10                      & level 7 \\
    \bottomrule
    \end{tabularx}
    \caption[Default Audio-Codec Settings]{Default audio-codec settings for the audio codecs used in our tests. The {\em Range} column shows an option's range while the {\em Setting} column displays the value used in testing. CBR stands for constant bit rate, VBR stands for variable bit rate.}
    \label{tbl:setup:dflt+codec+settings}
\end{table}

Instead of finding and using optimal parameters for each audio codec, presets were selected to represent frequently used default values. Our study prioritizes ecological validity and reproducibility over audio-codec tweaking when evaluating real-world audio-codec behavior under general use setups. Therefore, rather than describing theoretical audio-codec restrictions, reported results describe real-world deployment conditions. \Cref{tbl:setup:dflt+codec+settings} provides an overview of the settings used for the various audio encoding formats.
\par

\subsection{Visualization Tools}
\label{sec:setup:visual}

\begin{figure*}
    \centering
    \subfloat[Spectrum]{%
        \includegraphics[trim=0 120 36 20,clip,width=0.345\textwidth]%
        {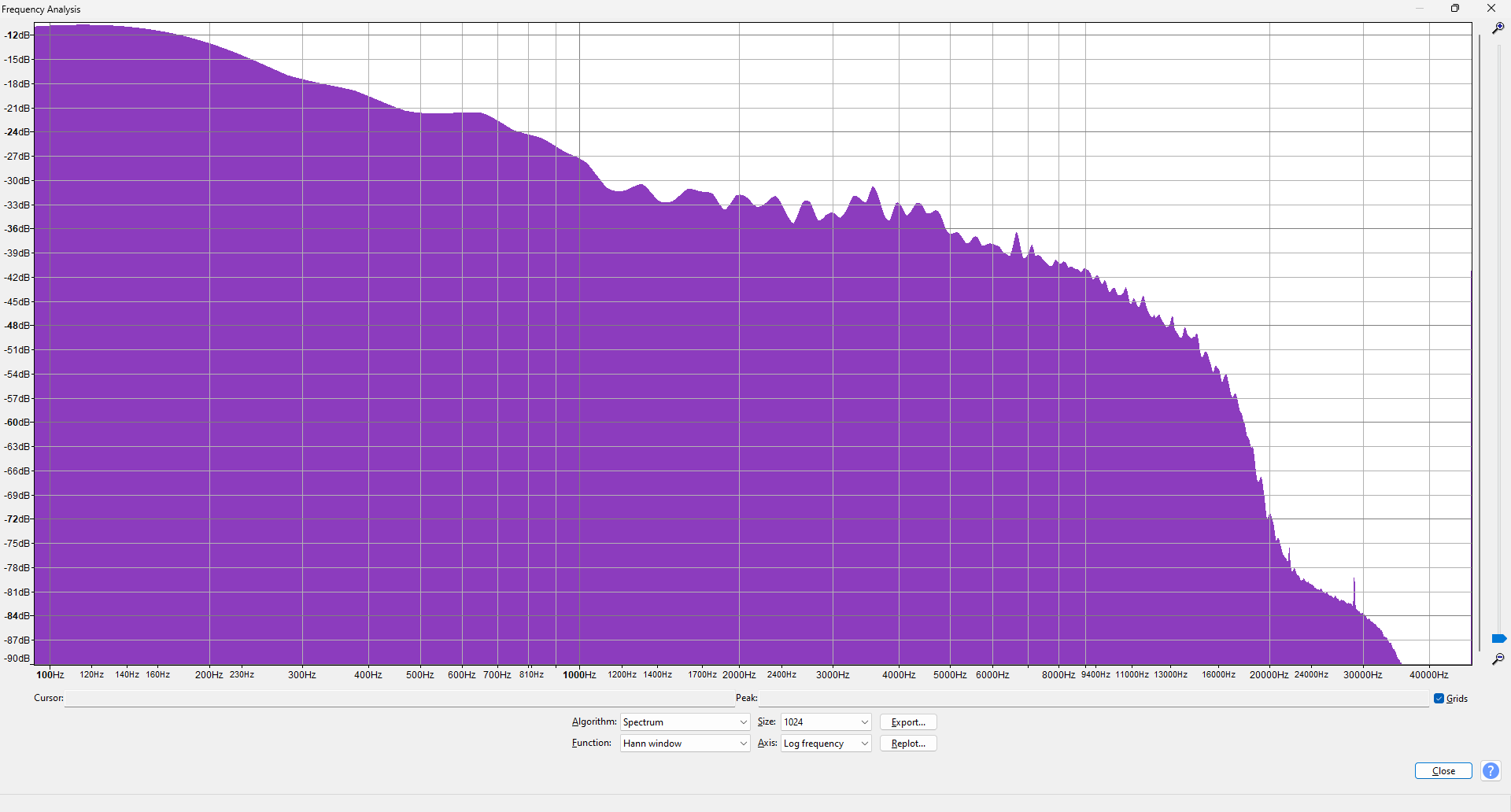}%
        \label{fig:setup:visual:spectrum+sample}
    }\subfloat[Spectrogram]{%
        \includegraphics[trim=5 5 10 95,clip,width=0.33\textwidth]%
        {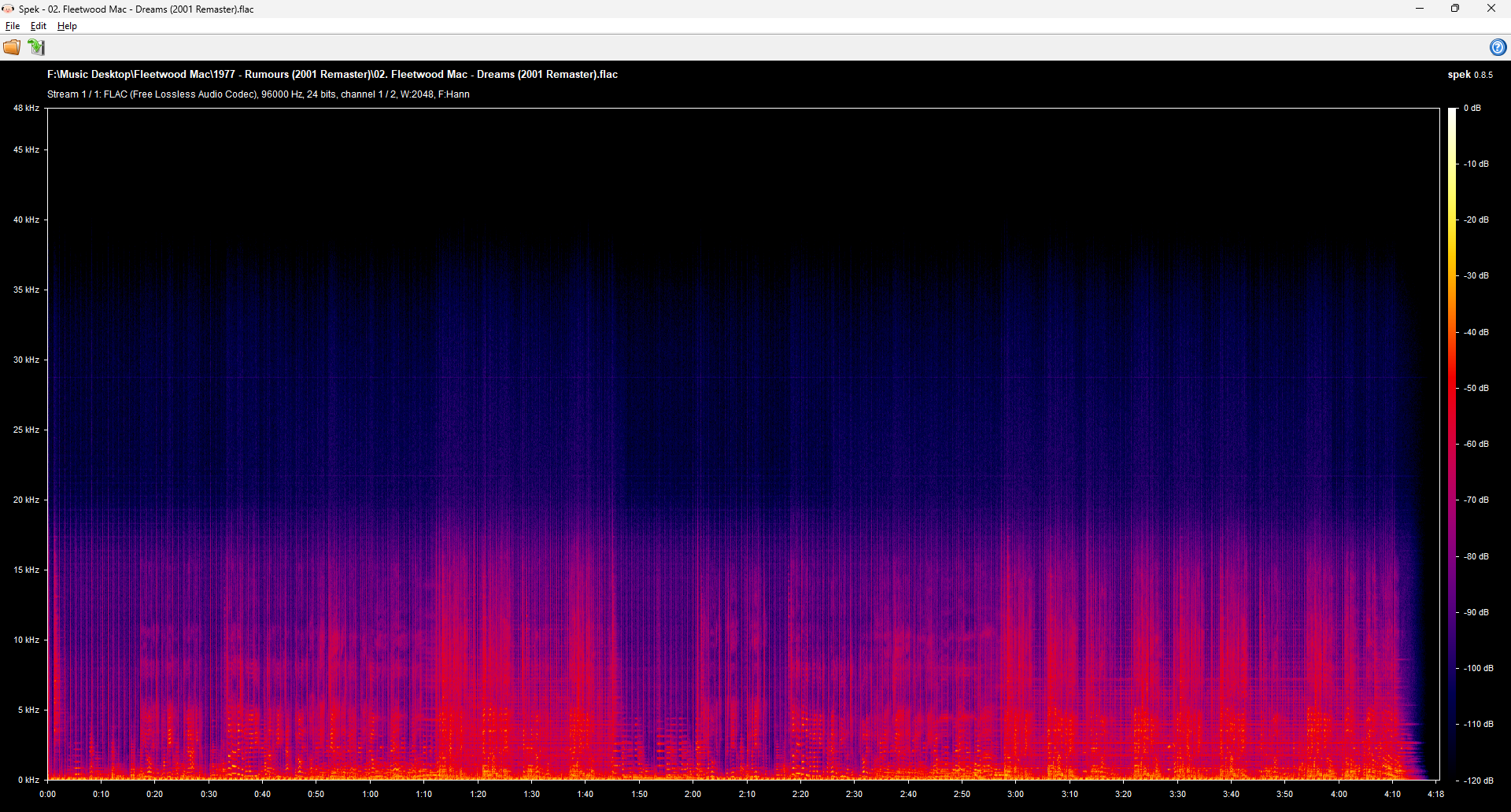}%
        \label{fig:setup:visual:spectrogram}
    }\subfloat[Sound field]{%
        \includegraphics[trim=40 114 40 118,clip,width=0.295\textwidth]%
        {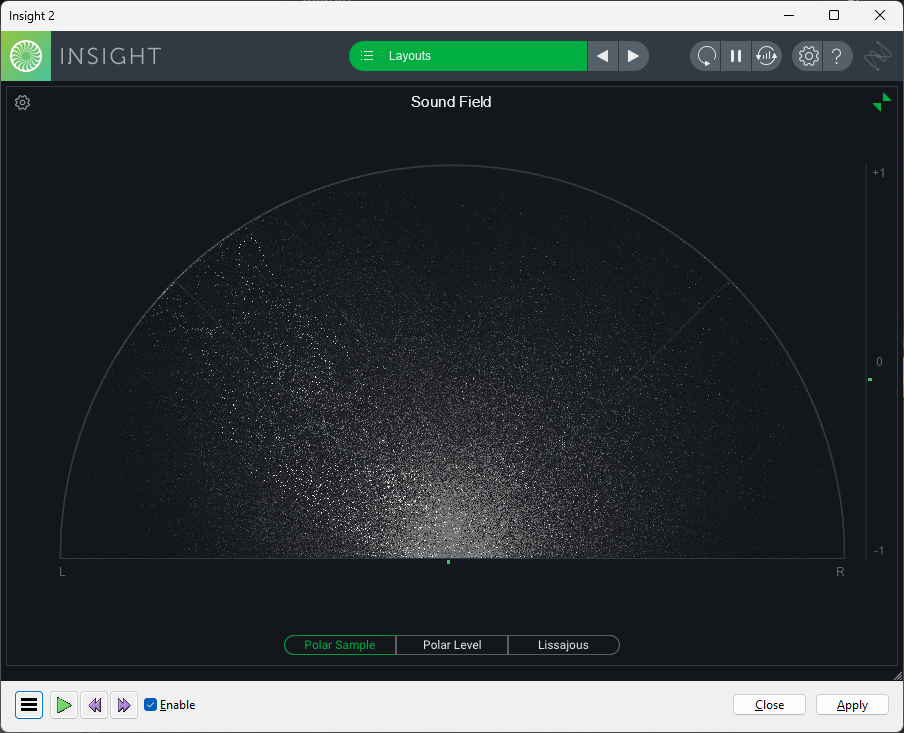}%
        \label{fig:setup:visual:sound+field}
    }
    \caption{Perceptual-audio quality visualization.
    \protect\subref{fig:setup:visual:spectrum+sample} Audacity generated spectrum shows peak loudness within the highest frequency (\KHZ{48}); x-axis shows time, y-axis shows frequency, colored spectrum indicates loudness in \DB{}. \protect\subref{fig:setup:visual:spectrogram} Spek generated spectrogram shows loudness range (\DB{0} to \DB{-120}) of an audio signal's frequency range; x-axis shows time, y-axis shows frequency range, colored spectrum indicates loudness in \DB{}.
    \protect\subref{fig:setup:visual:sound+field} Sound-field visualization generated by Audacity using the {\em Insight 2} plug-in shows width, depth, and height of the sound stage as well as the location of individual sound sources within that sound stage. The half-circle line indicates sound-stage front while bright dots indicate stereo images.
    Visualizations based on {\em Dreams} by {\em Fleetwood Mac} (2001 remaster vinyl quality, sampled at \KHZ{96}).}
    \label{fig:setup:visual:all}
\end{figure*}

To visualize perceptual quality a combination of spectral analysis tools were employed. These include the built-in spectrum generator of \textcite{AudacityWebsite} (\cf\ \cref{fig:setup:visual:spectrum+sample}), spectrogram analysis by \textcite{SpekWebsite} (\cf \cref{fig:setup:visual:spectrogram}), and sound-field visualizations provided by \textcite{IzotopeWebsite} (\cf \cref{fig:setup:visual:sound+field}). Each type of visualization represents a different aspect of aural perceptual quality. Spectrum analysis shows how audio-compression techniques affect the loudness of a digital-audio signal through its frequency response. Spectrograms display time, frequency, and amplitude with a color range for easy visual differentiation. Sound fields reveal how audio-compression techniques affect the {\em stereo image} and the {\em sound stage}, \ie how audio signals are distributed between the left and the right channel of a stereo system and the width, depth, and height of an audio signal, respectively.
\par

\subsection{Perceptual Evaluation of Audio Quality}
\label{sec:setup:peaq}

\begin{table}
    \centering\sffamily\smaller
    \begin{tabularx}{0.65\linewidth}{L{1.5}R{0.5}}
        \toprule
        Impairment                     & ODG \\
        \midrule
        Imperceptible                  &  0 \\
        Perceptible, not annoying      & -1 \\
        Slightly annoying              & -2 \\
        Annoying                       & -3 \\
        Very annoying                  & -4 \\
        \bottomrule
    \end{tabularx}
    \caption[Objective Difference Grade (ODG)]{{\em Objective Difference Grade} (ODG) scoring range, after \textcite{ITU:BS.1387-2:2023}. A generated score from the objective measurement corresponds to a {\em Subjective Difference Grade} (left column) qualitatively expressed as {\em Impairment}. ODG score resolution is limited to one decimal.}
    \label{tbl:setup:peaq:odg}
\end{table}

To quantify the impact of audio-compression techniques on digital audio signals an objective scoring system is required. {\em Perceptual Evaluation of Audio Quality} (PEAQ) \autocite{Thiede:PEAQITUSOMPAQ:2000} is an ITU-standardized algorithm to assess the perceptual quality of digital audio. PEAQ generates an {\em Objective Difference Grade} (ODG) score (\cf\ \cref{tbl:setup:peaq:odg}), expressed in terms of a quality gap between a tested audio signal and its respective original audio signal.
\par

There are two versions of PEAQ: basic PEAQ (BPEAQ) and advanced PEAQ (APEAQ) \autocite{Kabal:EIITUPEAQ:2002}. Both use the same {\em Fast Fourier Transform} (FFT) model to convert time-domain signals into frequency-domain signals for evaluation. APEAQ additionally includes a {\em filter-bank ear model} to accurately simulate the human auditory system's cochlear response. \Textcite{Delgado:CWSUP:2020} report BPEAQ to be less computational intensive than APEAQ, while APEAQ is the more appropriate version of PEAQ for newer audio codecs such as AAC and Vorbis. This is due to these audio codecs' perceptual models being more complex. BPEAQ's model output variable (MOV) {\em total noise-to-mask ratio} (\TNMR) is well-correlated with subjective listening tests \autocite{Dick:ODAQ:2025}. \TNMR is measured in negative decibels and ideally should be as far as possible from \DB{0} while at the same time \TNMR is trending close to \DB{0}. In our study, we calculate both basic and advanced PEAQ's ODG scores as well as basic PEAQ's \TNMR value to showcase the performance of each scoring metric.
\par

Additional models exist, which improve on PEAQ based on past evaluation. Of these the {\em 2f} model \autocite{Kastner:AEMESQ:2019} performs significantly better than others \autocite{Torcoli:OMPAQR:2021}. The 2f model uses the \CMD{ADB} and the \CMD{AvgModDiff1} MOVs of the BPEAQ model to calculate a {\em Mean MUSHRA Score Estimate} (\MMSEST) \autocite{Kastner:AEMESQ:2019} as shown in \cref{eq:setup:codec:MMSest+formula}.
\begin{equation}
    \begin{split}
        \MMSEST= & \,\frac{49.73}{1+(-0.0315\cdot\text{AvgModDiff1}-0.73)^2}\\
                 & -46.96\cdot\text{ADB}+147.12\\
    \end{split}
    \label{eq:setup:codec:MMSest+formula}
\end{equation}
The result of \cref{eq:setup:codec:MMSest+formula} is then clamped to the range [0\ldots 100] to calculate the actual 2f score as shown in \cref{eq:setup:codec:2f+formula}.
\begin{equation}
    \text{2f score}=\min(100,\max(0,\MMSEST))
    \label{eq:setup:codec:2f+formula}
\end{equation}
Past experiments have shown that the 2f model's average score is the closest equivalent to a subjective listening test result \autocite{Kastner:AEMESQ:2019}.
\par

We use McGill University's PEAQ Matlab model \autocite{Kabal:EIITUPEAQ:2002} to calculate a BPEAQ ODG score and extract the \TNMR MOV. In addition, \CMD{ADB} and \CMD{AvgModDiff1} MOVs are extracted to calculate a 2f score. To calculate APEAQ's ODG score, \CMD{GstPEAQ} \autocite{Holters:GSTPEAQ:2015} is utilized. We note that \CMD{GstPEAQ} is an open-source project that does not fully conform to \textcite{ITU:BS.1387-2:2023}. However, pre-study experiments have shown that \CMD{GstPEAQ} performs within reasonably close limit, \ie 0.03 difference in root mean square error, to \textcite{ITU:BS.1387-2:2023} and is therefore expected to perform similar to a subjective listening test \autocite{Holters:GSTPEAQ:2015}.
\par

\subsection{Test Samples}
\label{sec:setup:samples}

We conducted our experiments with 50{\small\textplus} digital-audio files, using high-fidelity conversions of CDs, DVDs, vinyls, and cassettes from a wide range of music genres and epochs. For each sample, we collected the file size, the sampling rate, and the original bit rate used for conversion. Each sample was then encoded using our chosen audio codecs at their respective quality settings (\cf\ \cref{tbl:setup:dflt+codec+settings}). To showcase the impact of audio-compression techniques on the perceptual quality of digital audio, in the following we chose a set of audio examples for visualizations that present high levels of perceptual quality as well as high degrees of accuracy in stereo imaging and sound-stage appearance.
\par

\section{Evaluation}
\label{sec:evaluation}

\subsection{Audio-Codec Performance}
\label{sec:evaluation:performance}

\begin{table}
    \newcommand{\DFM}{{\em Dreams} Fleetwood Mac}
    \newcommand{\LMD}{{\em La mer II. Jeux de Vagues} Claude Debussy}
    \newcommand{\FP}{{\em Foreigner} Pallbearer}
    \newcommand{\SEWF}{{\em September} Earth, Wind \& Fire}
    \centering\sffamily\smaller
    \begin{tabularx}{\linewidth}{L{1.2}L{.7}L{1}R{1.2}R{.95}R{.95}}
        \toprule
        \mrow{2}{=}{Track} &
        \mrow{2}{=}{Codec} &
        \mrow{2}{=}{Setting} &
        \mrow{2}{=}{Compression Ratio} &
        \mrow{2}{=}{Encoding Speed} &
        \mrow{2}{=}{Decoding Speed} \\
         & & & & & \\
        \midrule
        \mrow{6}{=}{\DFM}  & FLAC   & level 6    & 68.17 & 37.91 & 117.93 \\
                           & MP3    & \kbps{320} & 89.59 &  5.92 & 158.79 \\
                           & MP3    & \kbps{128} & 95.84 &  7.20 & 210.69 \\
                           & AAC    & \kbps{256} & 91.57 &  8.86 &  37.86 \\
                           & AAC    & level 5    & 89.50 &  6.83 &  32.99 \\
                           & Vorbis & level 7    & 92.50 &  9.47 &  58.46 \\
        \midrule
        \mrow{6}{=}{\LMD}  & FLAC   & level 6    & 66.67 & 38.87 & 118.76 \\
                           & MP3    & \kbps{320} & 77.32 &  3.14 &  75.12 \\
                           & MP3    & \kbps{128} & 90.93 &  3.67 &  99.48 \\
                           & AAC    & \kbps{256} & 81.75 &  5.43 &  28.81 \\
                           & AAC    & level 5    & 84.16 &  5.01 &  29.26 \\
                           & Vorbis & level 7    & 87.77 &  9.09 &  46.43 \\
        \midrule
        \mrow{6}{=}{\FP}   & FLAC   & level 6    & 42.48 & 36.41 & 104.45 \\
                           & MP3    & \kbps{320} & 77.33 &  3.64 &  75.80 \\
                           & MP3    & \kbps{128} & 90.93 &  3.91 &  96.80 \\
                           & AAC    & \kbps{256} & 81.76 &  5.58 &  28.43 \\
                           & AAC    & level 5    & 84.41 &  5.41 &  29.33 \\
                           & Vorbis & level 7    & 86.70 &  8.36 &  44.04 \\
        \midrule
        \mrow{6}{=}{\SEWF} & FLAC   & level 6    & 29.98 & 26.33 &  82.76 \\
                           & MP3    & \kbps{320} & 93.06 &  6.02 & 148.92 \\
                           & MP3    & \kbps{128} & 97.22 &  8.10 & 207.89 \\
                           & AAC    & \kbps{256} & 94.38 &  9.06 &  38.02 \\
                           & AAC    & level 5    & 92.51 &  6.64 &  34.07 \\
                           & Vorbis & level 7    & 94.48 &  9.09 &  56.52 \\
        \bottomrule
    \end{tabularx}
    \caption[Dataset Audio Codec Compression Ratios]{Excerpt from our codec-performance dataset showing audio-codec compression as percentage ratio and encoding as well as decoding speeds in samples per microsecond using the audio-codec configurations established in \cref{tbl:setup:dflt+codec+settings}.}
    \label{tbl:evaluation:codec+compression+dataset}
\end{table}

\Cref{tbl:evaluation:codec+compression+dataset} provides excerpts from our dataset for selected entries along with the chosen quality setting, compression ratio as a percentage, and encoding speed as well as decoding speed in samples per microsecond. The full dataset, \textcite{FMediaACCDGithubRepo}, contains 50{\small\textplus} entries. Each entry specifies additional metadata (\eg genre) as well as original file size, number of digital samples, compression bit rate, compressed file size, original encoding algorithm, and encoding time as well as speed for encoding and decoding for each of the tested audio codecs.
\par

\begin{table}
    \newcommand{\hl}[1]{\bfseries #1}
    \centering\sffamily\smaller
    \begin{tabularx}{\linewidth}{L{1}L{.7}L{1.15}R{1.25}R{.95}R{.95}}
        \toprule
         &
        \mrow{2}{=}{Codec} &
        \mrow{2}{=}{Setting} &
        \mrow{2}{=}{Compression Ratio} &
        \mrow{2}{=}{Encoding Speed} &
        \mrow{2}{=}{Decoding Speed} \\
         & & & & & \\
        \midrule
        Lossless           & FLAC   & level 6    &     38.74  & \hl{31.92} &      92.47  \\
        \midrule
        \mrow{5}{=}{Lossy} & MP3    & \kbps{320} &     83.99  &      4.90  &     112.30  \\
                           & MP3    & \kbps{128} & \hl{92.44} &      5.70  & \hl{148.93} \\
                           & AAC    & \kbps{256} &     86.73  &      7.12  &      33.07  \\
                           & AAC    & level 5    &     87.42  &      6.03  &      31.97  \\
                           & Vorbis & level 7    &     88.68  &      8.69  &      46.39  \\
        \bottomrule
    \end{tabularx}
    \caption[Average Audio Codec Compression Ratios]{Average audio-codec compression as percent ratio and encoding as well as decoding speeds in samples per microsecond. Best performance in each category has been \hl{marked}.}
    \label{tbl:evaluation:codec+compression+speed}
\end{table}

\Cref{tbl:evaluation:codec+compression+speed} presents average values for compression ratio, encoding speed and decoding speed of our dataset. FLAC, the only lossless compression scheme tested, achieves an average encoding speed of 32 samples per microsecond, \ie 32 million samples per second. MP3 at \kbps{128}, a lossy compression scheme, exhibits both the best average compression ratio at 92\% as well as the best decoding speed at 149 samples per microsecond, \ie 149 million samples per second.
\par

\subsection{Perceptual Quality}
\label{sec:evaluation:quality}

\begin{figure}
    \includegraphics[width=\linewidth]{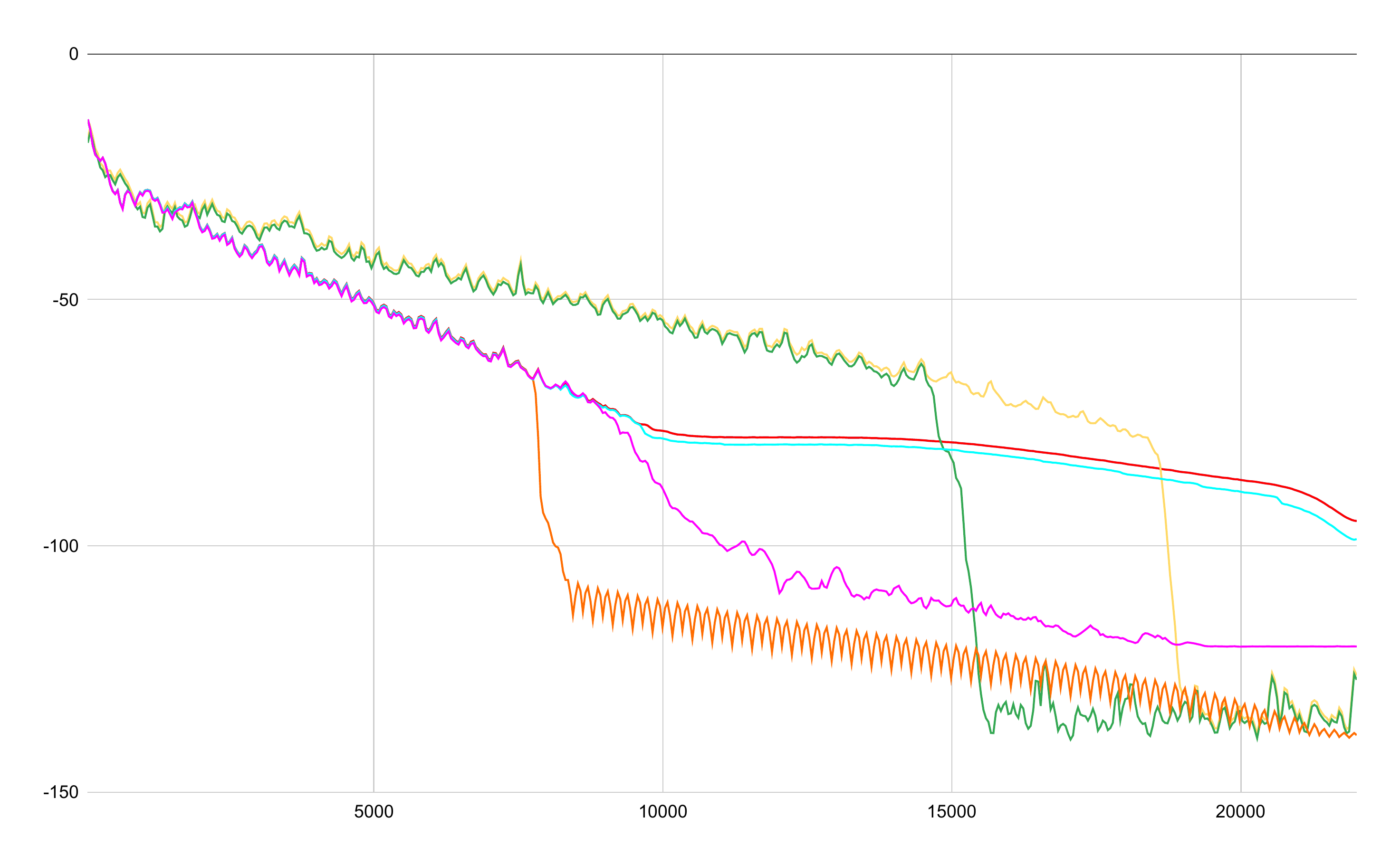}%
    \caption[Spectra of Tested Audio Codecs]{Spectra for all tested audio codecs (exported from Audacity). X-axis denotes frequency range up to \KHZ{24}. Y-axis shows loudness range from \DB{0} to \DB{-150}. Color-coded graphs refer to uncompressed (black), FLAC level 6 (red), MP3 CBR \kbps{320} (yellow), MP3 CBR \kbps{128} (green), AAC CBR \kbps{256} (orange), AAC VBR level 5 (teal), and Vorbis VBR level 7 (pink). Graphs represent encoded audio-signal outputs for the respective audio codec. Note the spectrum for the uncompressed audio signal is completely aligned with that of FLAC level 6, meaning both exhibit the exact same loudness across the entire frequency range. Spectra of both MP3, AAC CBR \kbps{256}, and Vorbis encoded audio diverge from uncompressed audio spectrum, revealing the impact of lossy compression schemes on the overall loudness.}
    \label{fig:evaluation:spectrum}
\end{figure}

\Cref{fig:evaluation:spectrum} visualizes the spectra of all tested audio codecs versus the uncompressed digital audio of {\em Iron Man} by Black Sabbath. The graph for FLAC level 6 aligns perfectly with the spectrum of the uncompressed audio signal, showing that the loudness of FLAC encoded audio exactly matches the loudness of the uncompressed audio across the entire spectrum. On the other hand, spectra for all audio encoded with lossy compression techniques deviate from the spectrum of the original audio signal. The same pattern of reduction in loudness can be seen for the interval \KHZ{15} to above \KHZ{20}. MP3 \kbps{320} and MP3 \kbps{128} elevate the loudness of the entire frequency range before cutoff occurs at \KHZ{18} and \KHZ{15}, respectively. AAC CBR \kbps{256} loudness reduction happens much earlier in the frequency range, around \KHZ{7}. To our surprise, AAC VBR level 5 loudness closely matches with the spectra for FLAC and uncompressed audio with only a slight loudness reduction at the high-frequency range. Vorbis level 7 also exhibits loudness reduction at high frequencies. However, unlike the other tested lossy compression techniques it manifests a smooth decline in loudness.
\par

\begin{figure}
    \centering
    \subfloat[][Uncompressed]{%
        \includegraphics[width=0.492\linewidth]{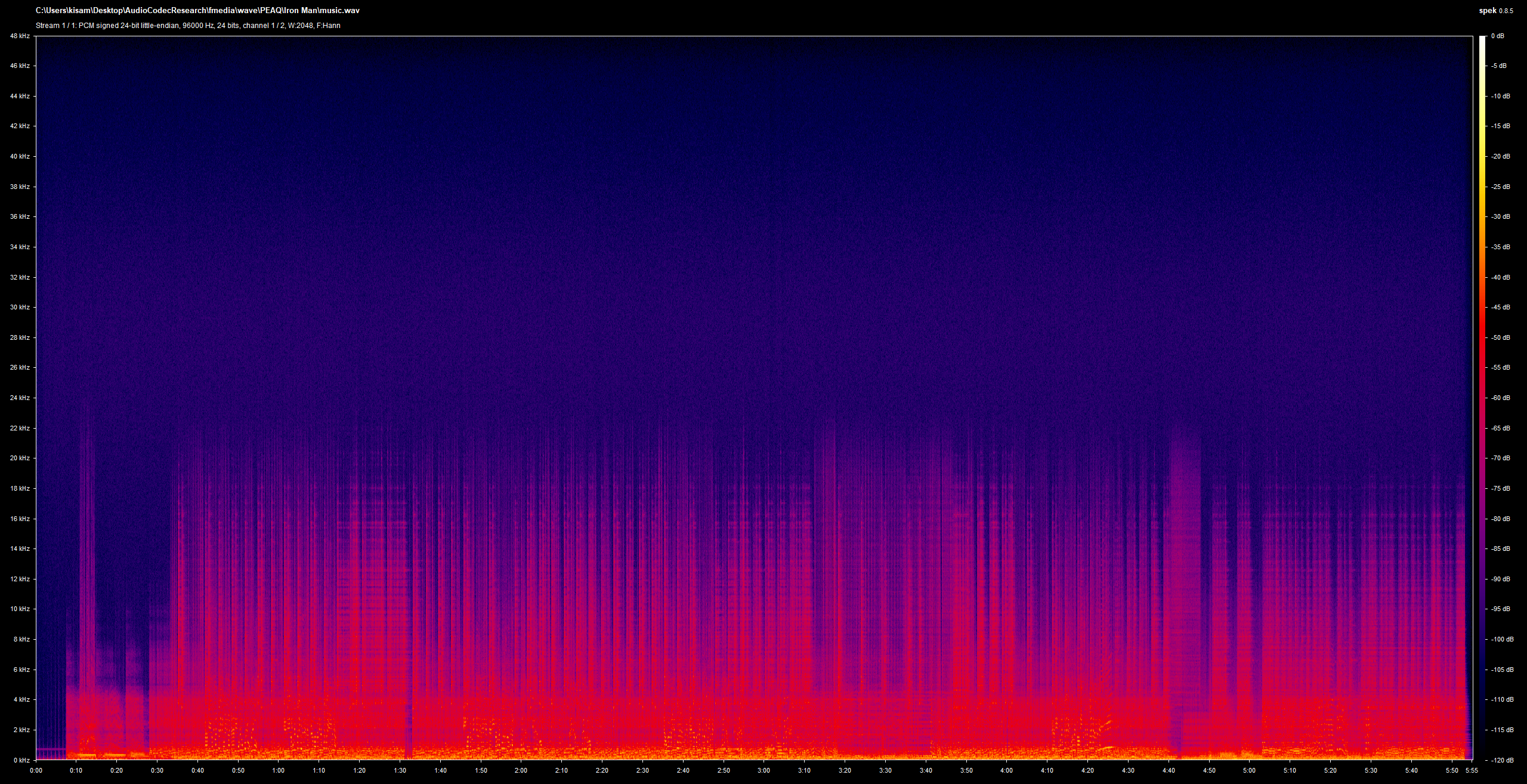}%
        \label{fig:evaluation:iron+man:a}
        \hspace*{0.492\linewidth}
    }\\
    \subfloat[][FLAC Level 6]{%
        \includegraphics[width=0.492\linewidth]{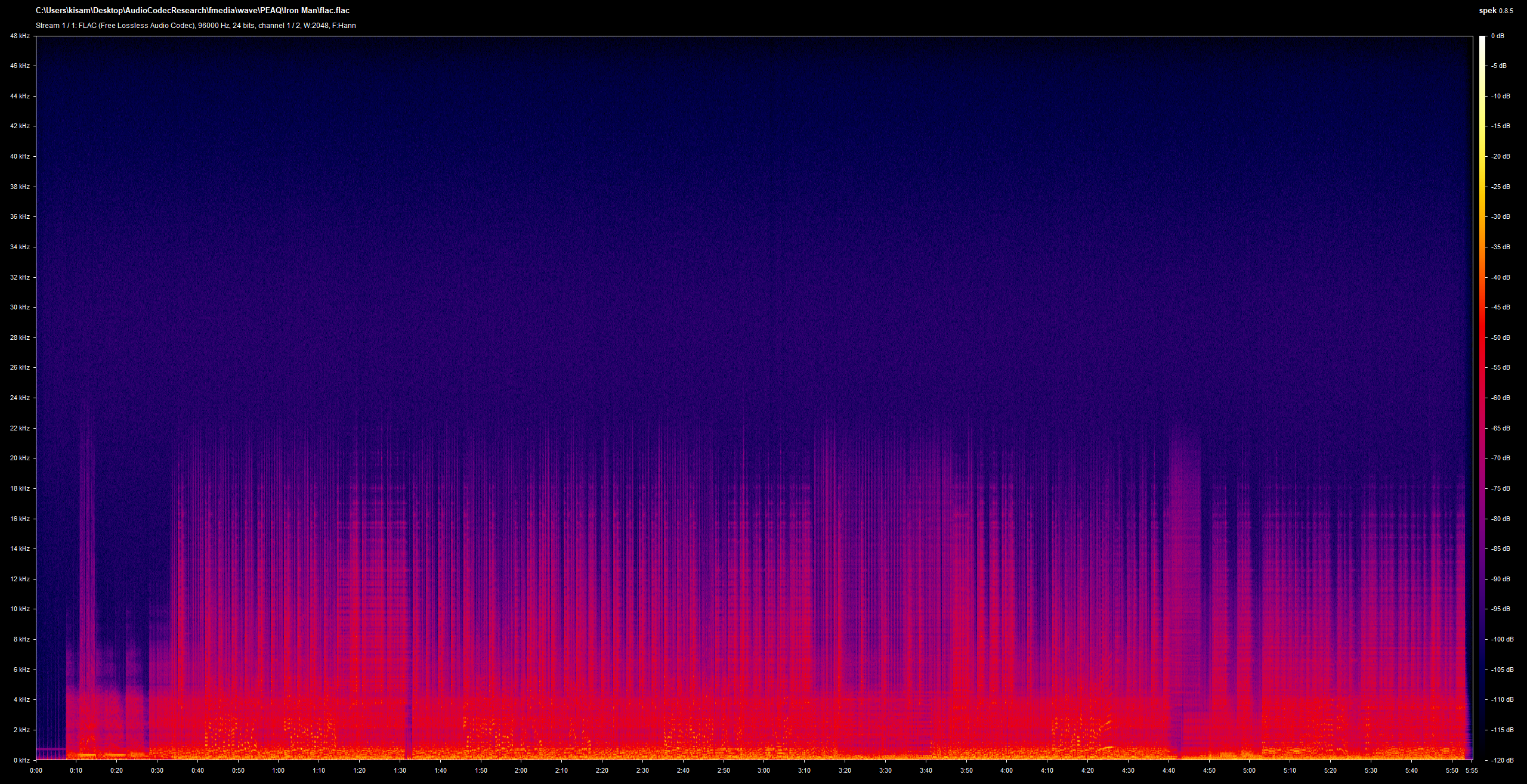}%
        \label{fig:evaluation:iron+man:b}
    }\subfloat[][MP3 CBR \kbps{320}]{%
        \includegraphics[width=0.492\linewidth]{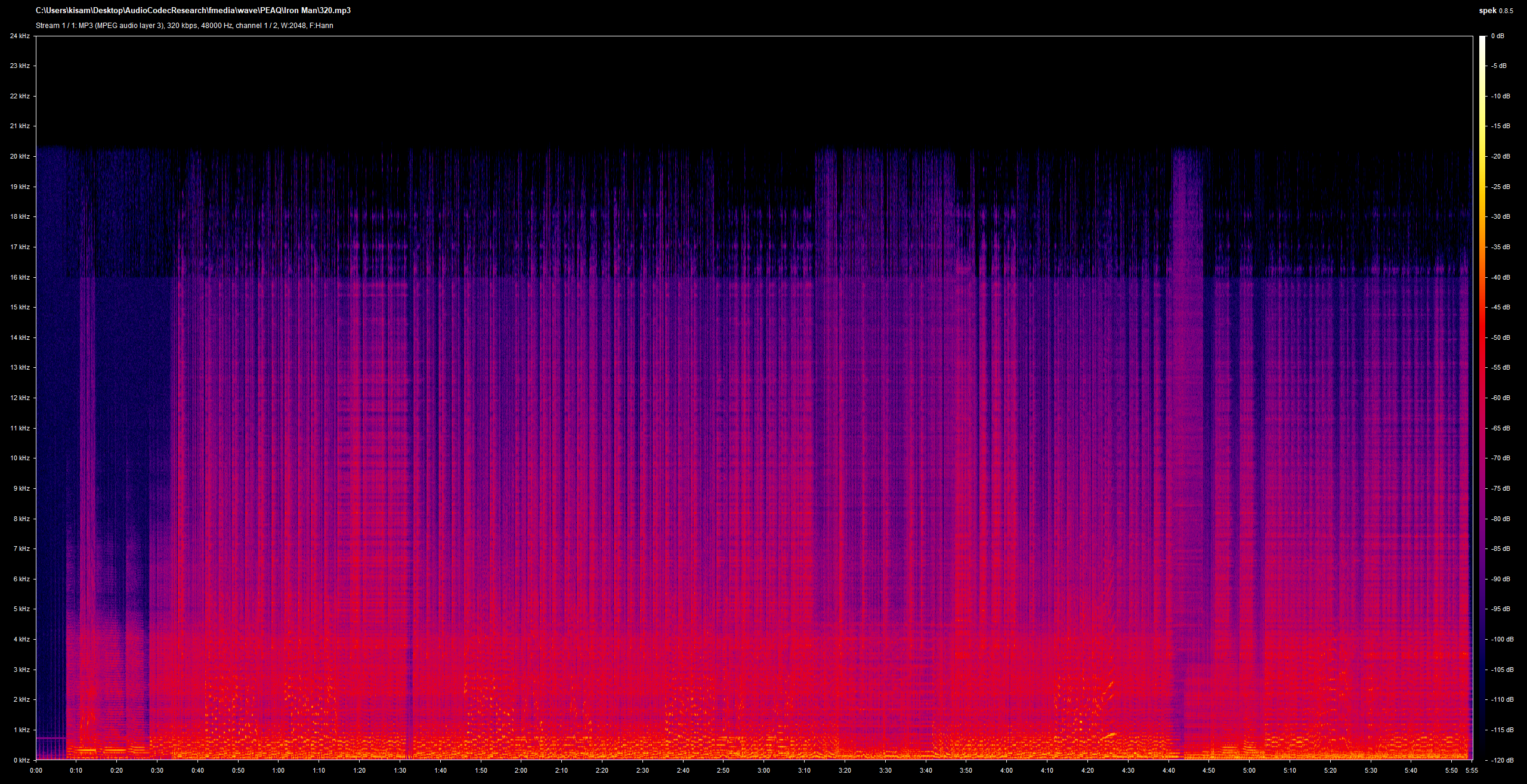}%
        \label{fig:evaluation:iron+man:c}
    }\\
    \subfloat[][MP3 CBR \kbps{128}]{%
        \includegraphics[width=0.492\linewidth]{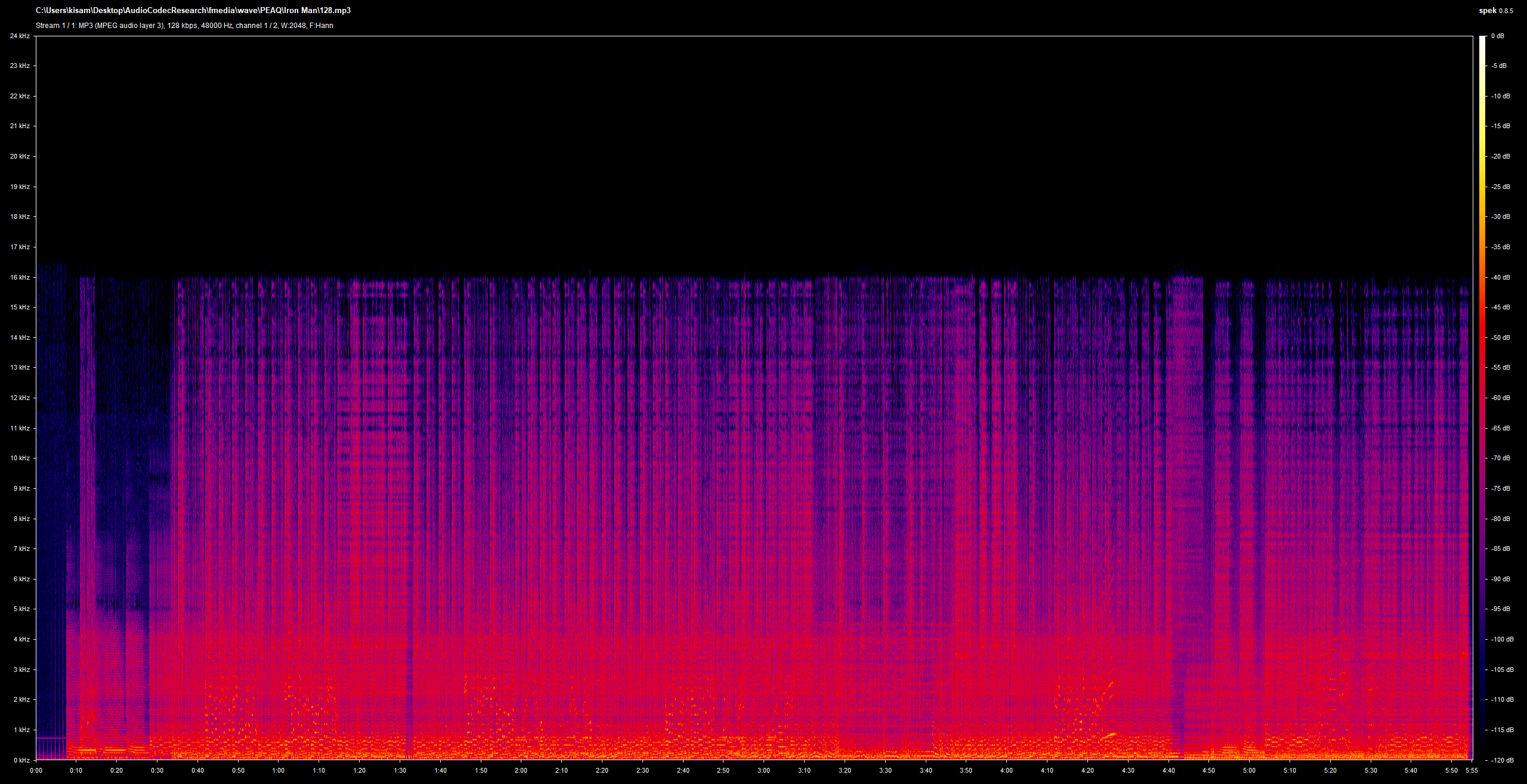}%
        \label{fig:evaluation:iron+man:d}
    }\subfloat[][AAC CBR \kbps{256}]{%
        \includegraphics[width=0.492\linewidth]{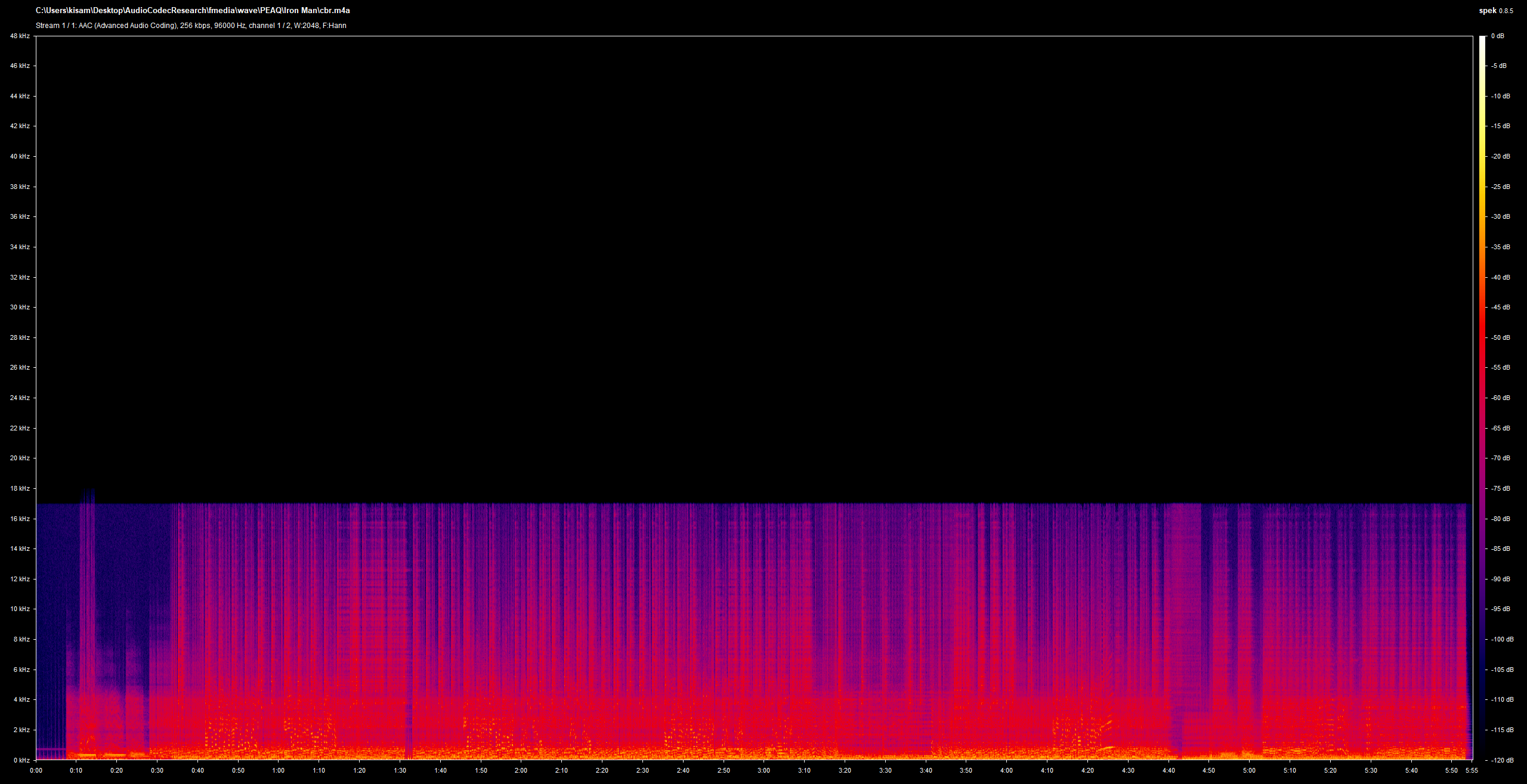}%
        \label{fig:evaluation:iron+man:e}
    }\\
    \subfloat[][AAC VBR level 5]{%
        \includegraphics[width=0.492\linewidth]{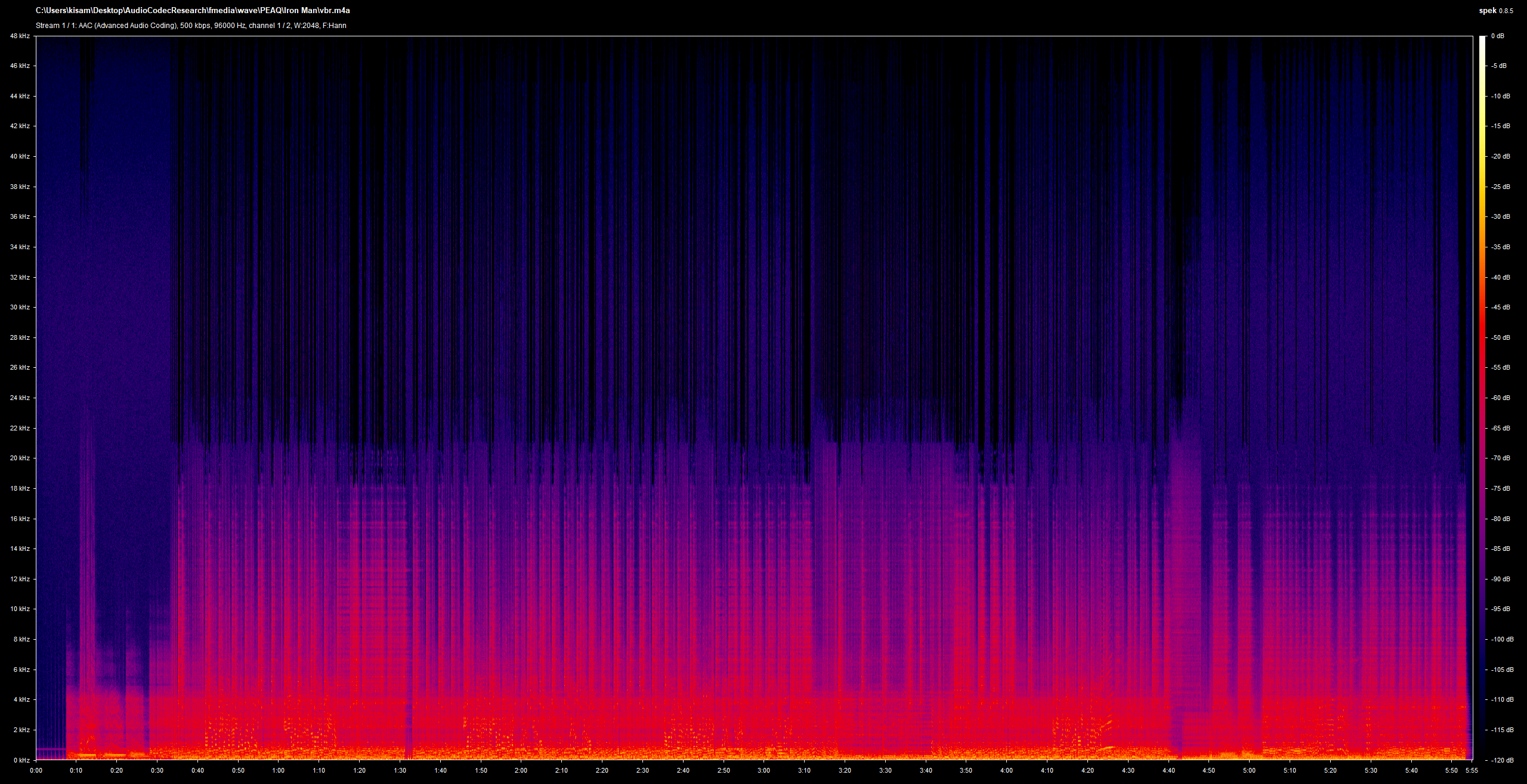}%
        \label{fig:evaluation:iron+man:f}
    }
    \subfloat[][Vorbis level 7]{%
        \includegraphics[width=0.492\linewidth]{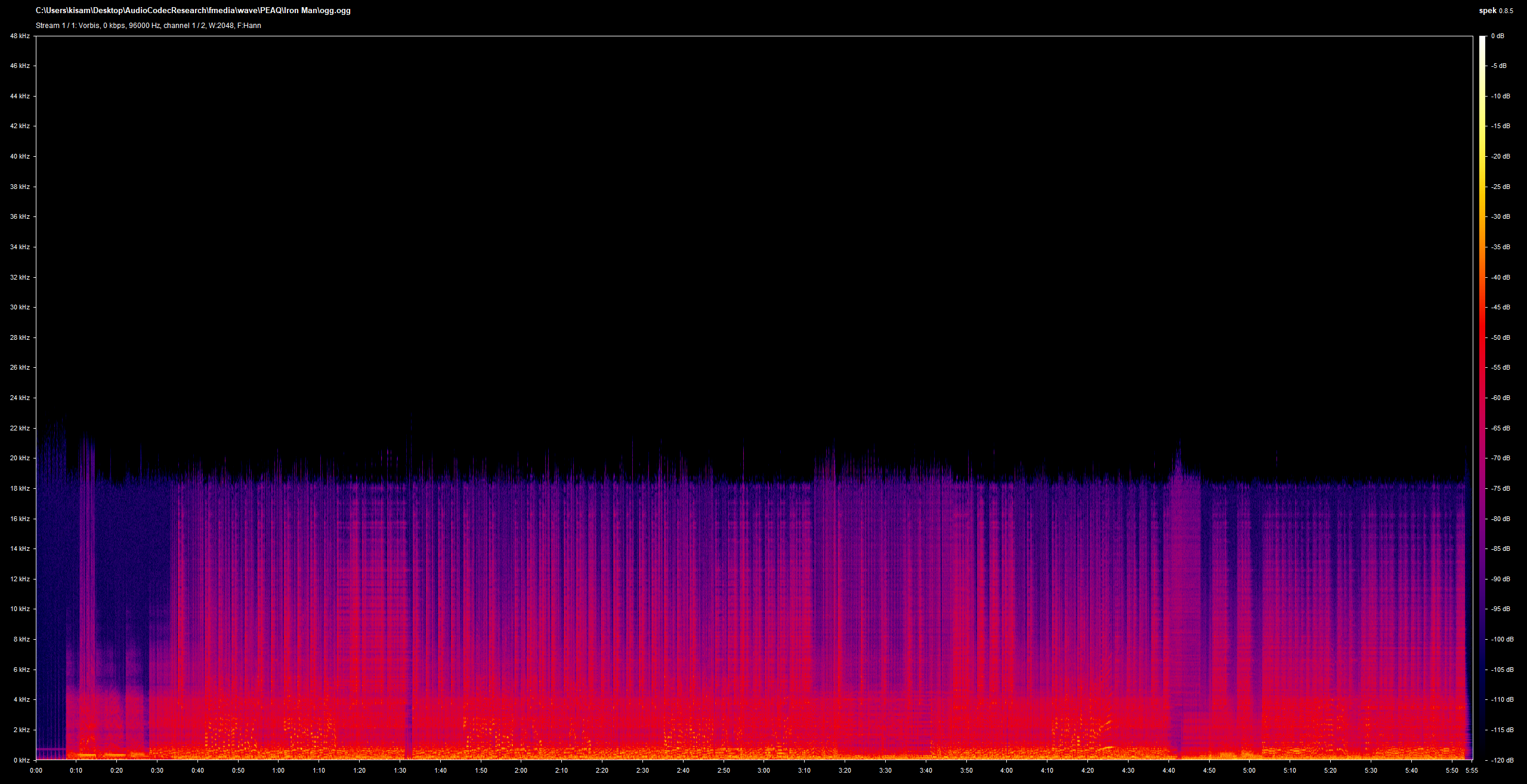}%
        \label{fig:evaluation:iron+man:g}
    }
    \caption[Audible Loudness Comparison ({\em Iron Man} by Black Sabbath)]{Audible loudness comparison for the track {\em Iron Man} by Black Sabbath using different encoding schemes. \protect\subref{fig:evaluation:iron+man:a} shows the spectrogram of the uncompressed signal while \protect\subref{fig:evaluation:iron+man:b}\,--\,\protect\subref{fig:evaluation:iron+man:g} show spectrograms for FLAC, MP3, AAC, and Vorbis, respectively. Note that MP3 requires a version of the track scaled down from \KHZ{96} to \KHZ{48}; the diagram for MP3 has been scaled in turn by a factor of two. All x-axes show duration of the tested audio signal while y-axes show the frequency range of the tested audio signal. Spectrograms indicate loudness in \DB{}.}
    \label{fig:evaluation:iron+man}
\end{figure}

\textcite{SpekWebsite} was used to further investigate and visualize the loudness change in encoded audio compared to uncompressed audio. \Cref{fig:evaluation:iron+man} shows that all audio encoded with lossy compression schemes exhibit very quiet loudness in the range \DB{-100} to \DB{-105}, marked dark blue in \cref{fig:evaluation:iron+man:a}, which is further reduced to near silence at \DB{-120}, marked black in \crefrange{fig:evaluation:iron+man:c}{fig:evaluation:iron+man:g}, for high frequencies above \KHZ{15}. MP3 at \kbps{320} maintains much of the original structure but exhibits slight attenuation in the upper frequencies, while MP3 at \kbps{128} shows a more pronounced high-frequency cutoff and reduced spectral detail. AAC and Vorbis demonstrate similar high-frequency constraints and reshape frequency-dependent energy in the mid-to-lower frequency. The redistribution of quantization noise is part of AAC's and Vorbis' perceptual encoding strategy to compress audio content while making sure the audio quality is perceptually masked \autocite{Chen:CPPGA:2005}.
\par

\begin{figure}
    \centering
    \subfloat[][Uncompressed]{%
        \includegraphics[width=0.492\linewidth]{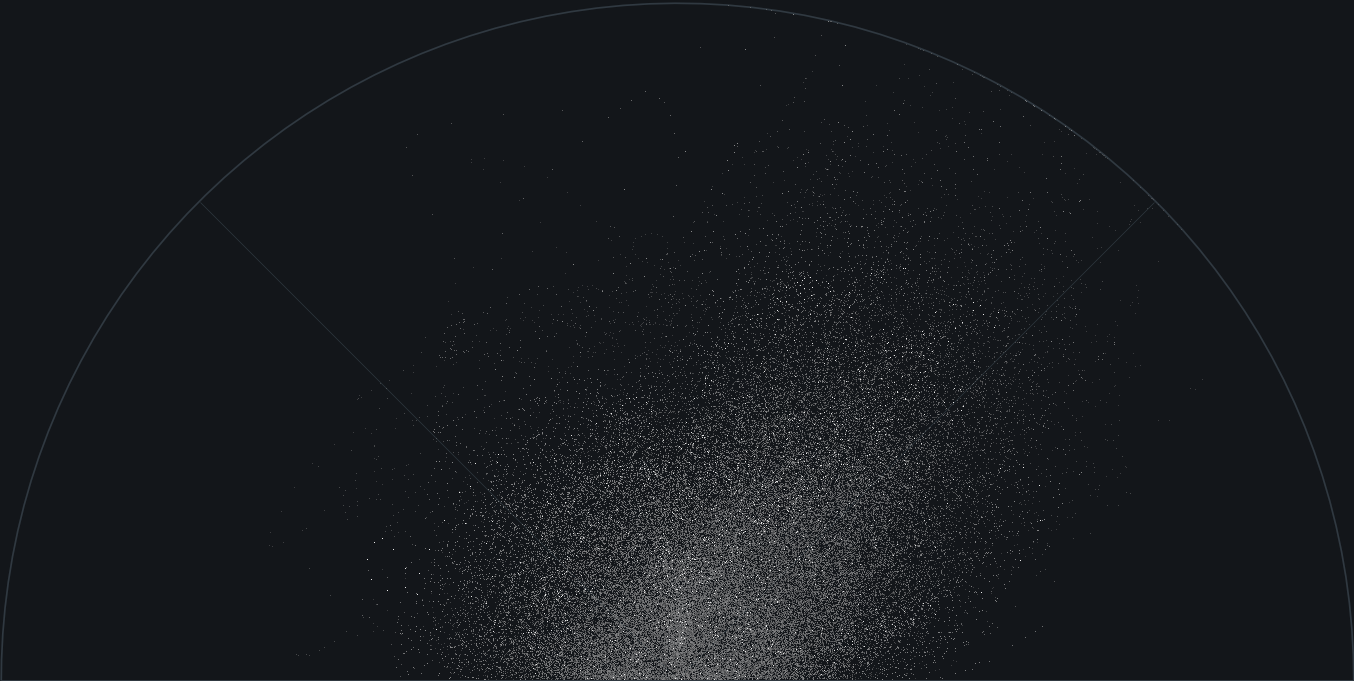}%
        \label{fig:evaluation:iron+man+sf:a}
        \hspace*{0.492\linewidth}
    }\\
    \subfloat[][FLAC level 6]{%
        \includegraphics[width=0.492\linewidth]{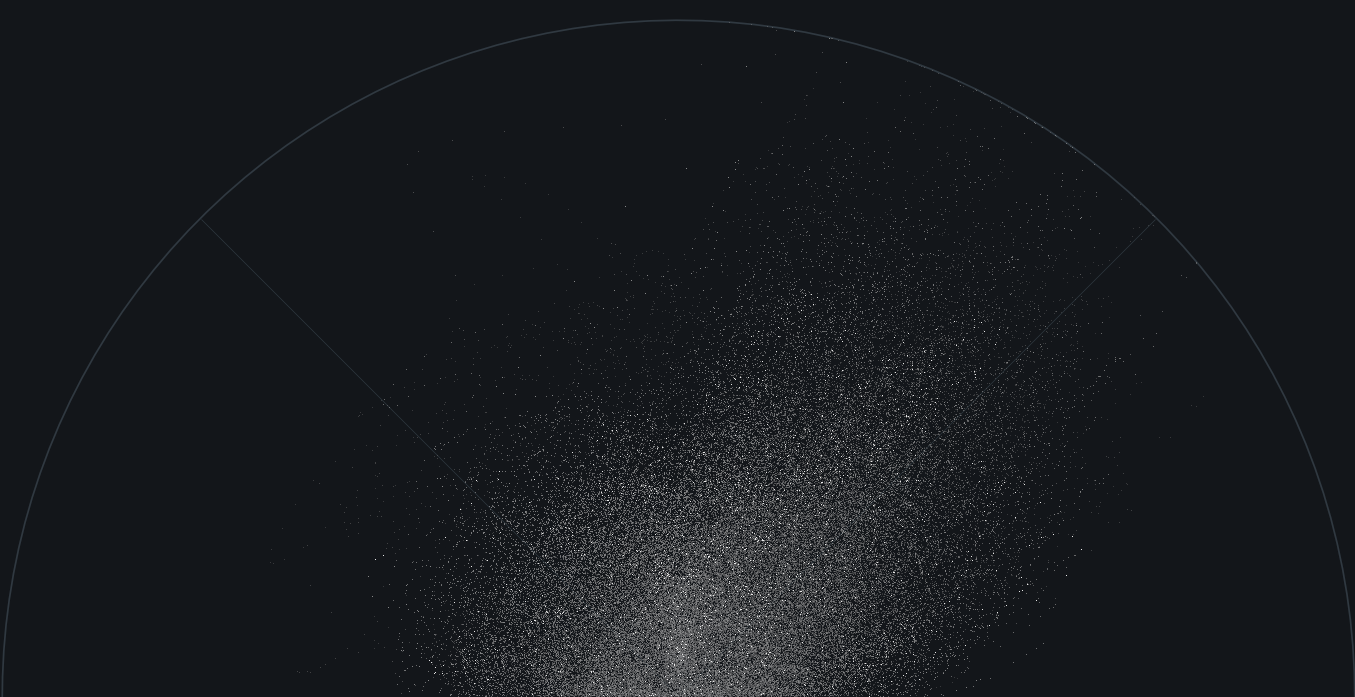}%
        \label{fig:evaluation:iron+man+sf:b}
    }\subfloat[][MP3 CBR \kbps{320}]{%
        \includegraphics[width=0.492\linewidth]{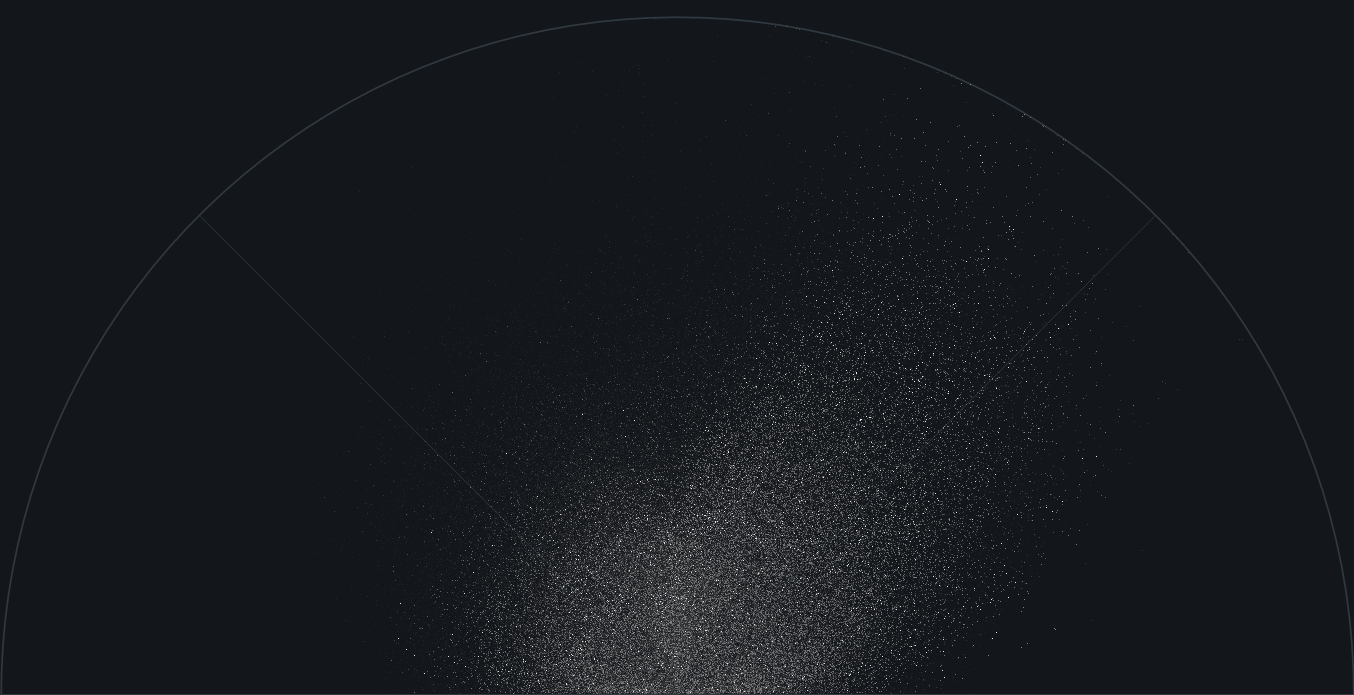}%
        \label{fig:evaluation:iron+man+sf:c}
    }\\
    \subfloat[][MP3 CBR \kbps{128}]{%
        \includegraphics[width=0.492\linewidth]{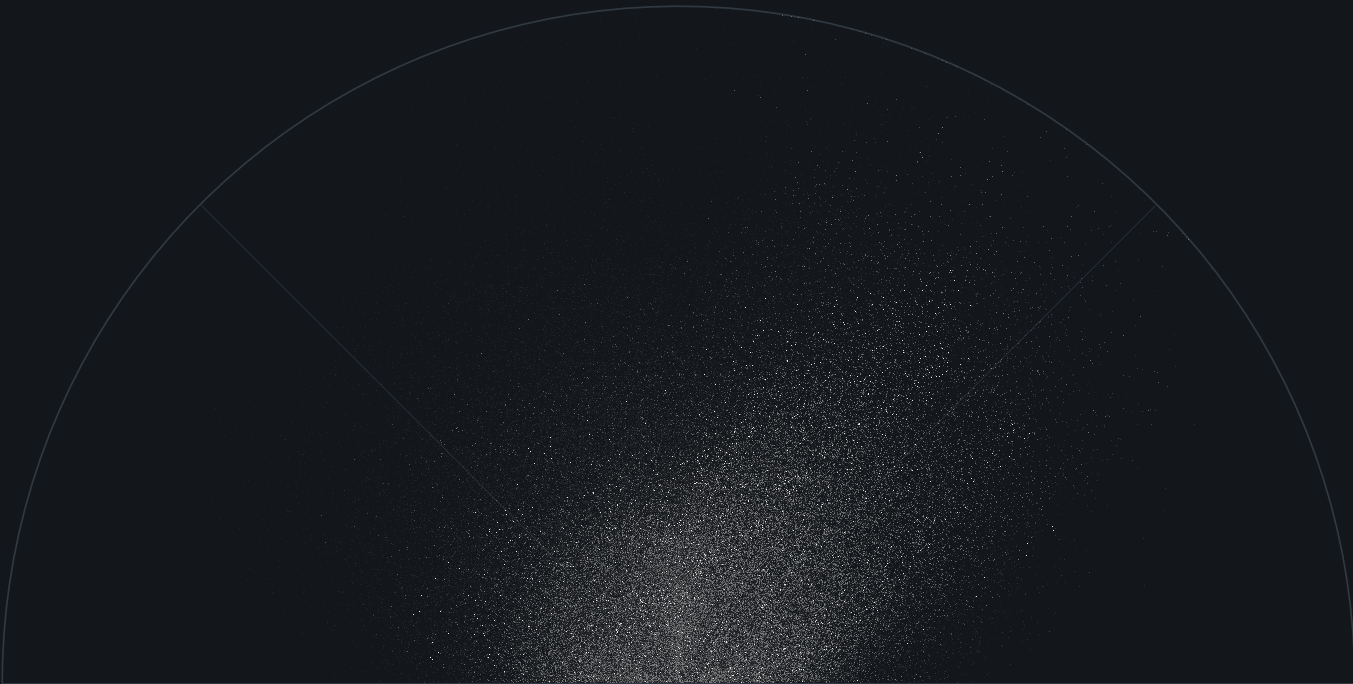}%
        \label{fig:evaluation:iron+man+sf:d}
    }\subfloat[][AAC CBR \kbps{256}]{%
        \includegraphics[width=0.492\linewidth]{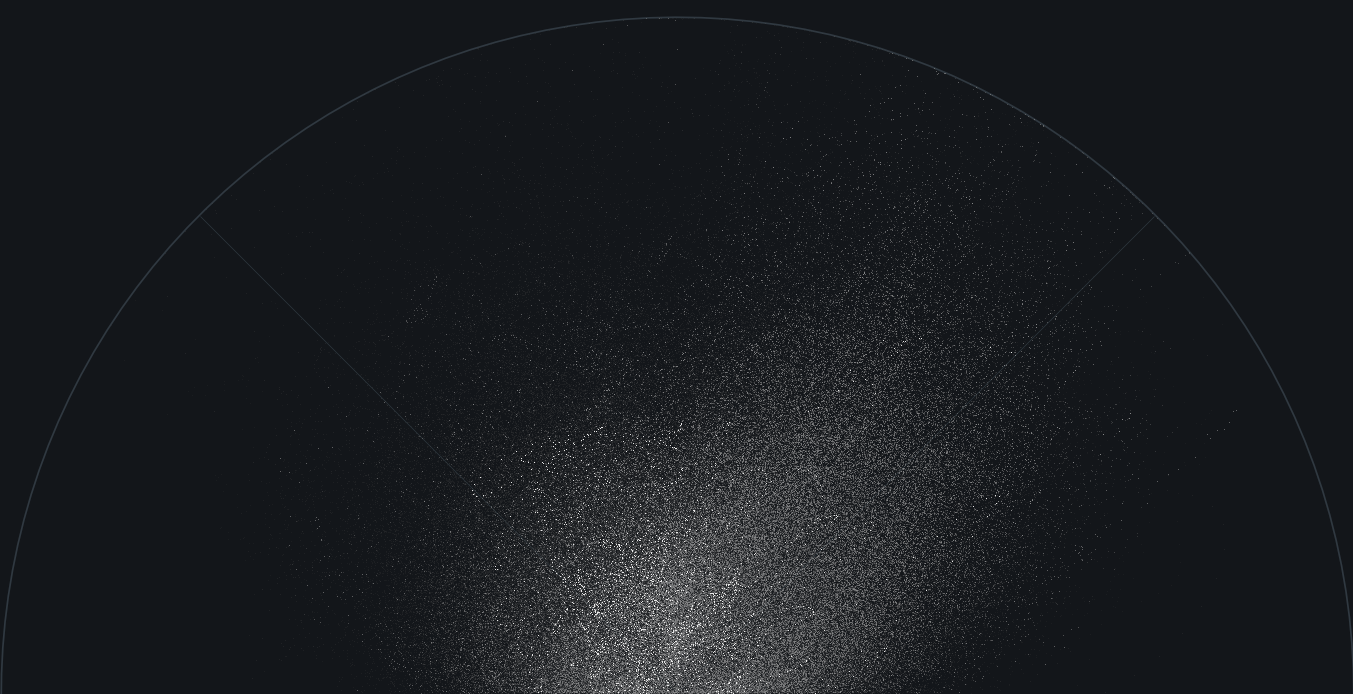}%
        \label{fig:evaluation:iron+man+sf:e}
    }\\
    \subfloat[][AAC VBR level 5]{%
        \includegraphics[width=0.492\linewidth]{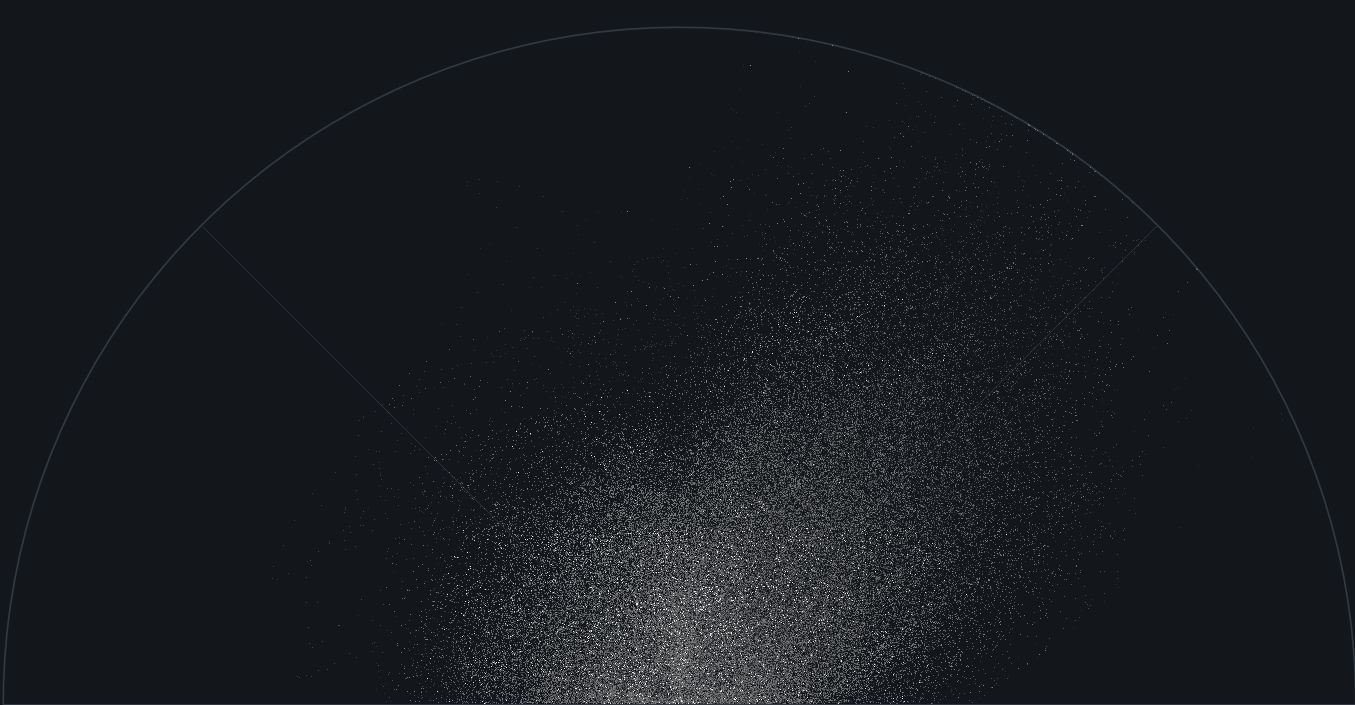}%
        \label{fig:evaluation:iron+man+sf:f}
    }\subfloat[][Vorbis VBR level 7]{%
        \includegraphics[width=0.492\linewidth]{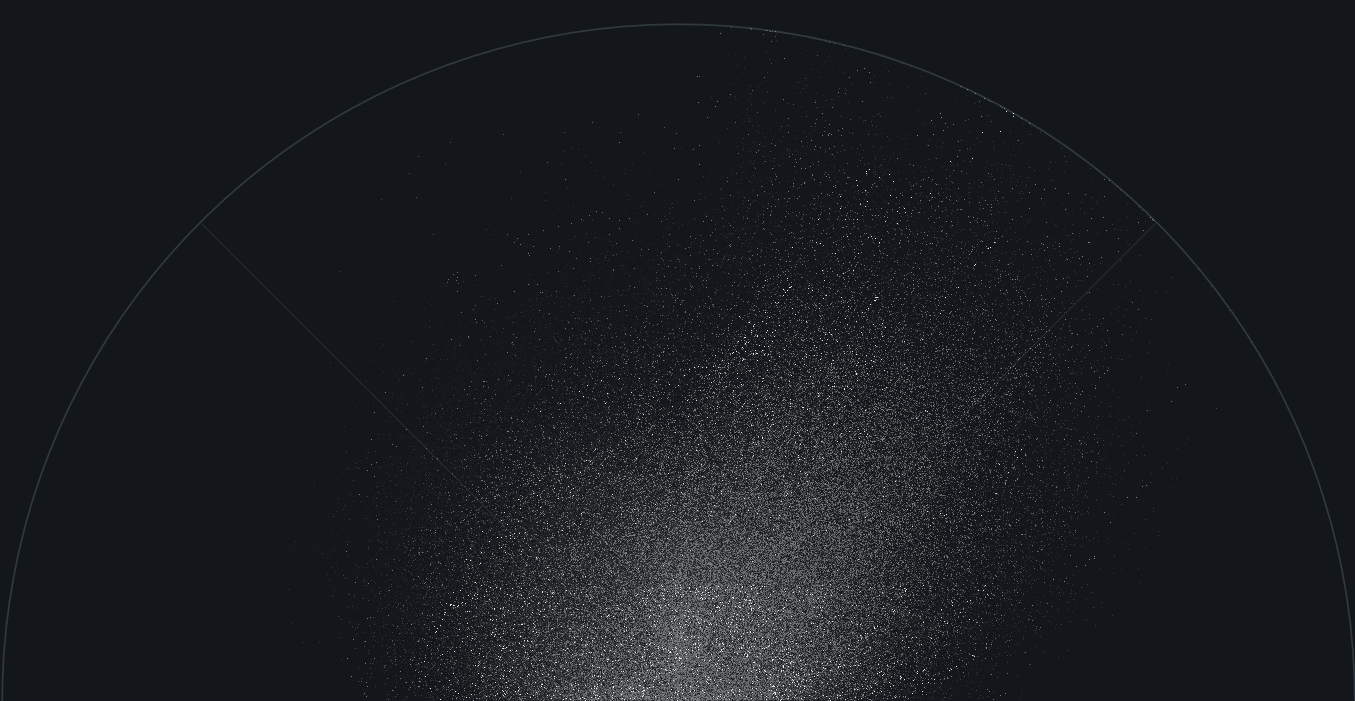}%
        \label{fig:evaluation:iron+man+sf:g}
    }
    \caption[Sound-Field Visualizations ({\em Iron Man} by Black Sabbath)]{Sound-field visualizations for the track {\em Iron Man} by Black Sabbath compressed using different audio encoding schemes. \protect\subref{fig:evaluation:iron+man+sf:a} shows the sound-field visualization from \MIN{3:00\,--\,3:30} for the uncompressed audio signal and \protect\subref{fig:evaluation:iron+man+sf:b}\,--\,\protect\subref{fig:evaluation:iron+man+sf:g} show the sound-field visualization for the same time interval encoded with FLAC, MP3, AAC, and Vorbis, respectively. Dot patterns should lean to the right, capturing the stereo image of the electric guitar's sound in the recording. Compared to the uncompressed audio signal's stereo image, FLAC encoded audio was able to accurately capture the stereo image while other lossy encoders exhibit some inaccuracies.}
    \label{fig:evaluation:iron+man+sf}
\end{figure}

\Cref{fig:evaluation:iron+man+sf} shows example sound-field visualizations for stereo images and sound stages capturing the \MIN{3:00\,--\,3:30} interval of {\em Iron Man} by Black Sabbath when Tony Iommi performs the guitar solo. Listening to the uncompressed audio the guitar solo appears slightly to the right, which leads to the expectation that the stereo image of this session should {\em lean} to the right of the center as well. As expected the sound field for the FLAC encoded audio in \cref{fig:evaluation:iron+man+sf:b} is nearly indistinguishable from that of the uncompressed audio. All the sound-field visualizations in \crefrange{fig:evaluation:iron+man+sf:b}{fig:evaluation:iron+man+sf:g} have their clusters lean in the expected direction. However, the clusters in \crefrange{fig:evaluation:iron+man+sf:c}{fig:evaluation:iron+man+sf:g}, representing audio encoded with lossy compression techniques, significantly deviate from the uncompressed sound field in \cref{fig:evaluation:iron+man+sf:a}. The sound-field clusters are too sparse for \crefrange{fig:evaluation:iron+man+sf:c}{fig:evaluation:iron+man+sf:d} using MP3 encoding, while the clustering for AAC CBR \kbps{256} in \cref{fig:evaluation:iron+man+sf:e} is too condensed in the center. However, sound-field clusters for both AAC VBR level 5 (\cf\ \cref{fig:evaluation:iron+man+sf:f}) and Vorbis level 7 (\cf\ \cref{fig:evaluation:iron+man+sf:g}) are very similar to the uncompressed audio sound field in \cref{fig:evaluation:iron+man+sf:a}.
\par

\begin{table}
    \newcommand{\DFM}{{\em Dreams} Fleetwood Mac}
    \newcommand{\LMD}{{\em La mer II. Jeux de Vagues} Claude Debussy}
    \newcommand{\FP}{{\em Foreigner} Pallbearer}
    \newcommand{\SEWF}{{\em September} Earth, Wind \& Fire}
    \centering\sffamily\smaller
    \begin{tabularx}{\linewidth}{L{1.3}L{.7}L{1.1}R{.9}R{.85}R{.9}R{1.25}}
        \toprule
        \mrow{2}{=}{Track} &
        \mrow{2}{=}{Codec} &
        \mrow{2}{=}{Setting} &
        \mrow{2}{=}{BPEAQ ODG} &
        \mrow{2}{=}{2f Score} &
        \mrow{2}{=}{APEAQ ODG} &
        \mrow{2}{=}{BPEAQ \TNMR} \\
         & & & & & & \\
        \midrule
        \mrow{6}{=}{\DFM}  & FLAC   & level 6    &  0.2 & 100 &  0.2 & -131.08 \\
                           & MP3    & \kbps{320} & -3.8 &  32 & -0.1 &   -0.77 \\
                           & MP3    & \kbps{128} & -3.8 &  32 & -0.1 &   -1.07 \\
                           & AAC    & \kbps{256} & -3.9 &  29 & -0.1 &    0.15 \\
                           & AAC    & level 5    & -3.9 &  29 & -0.1 &    0.15 \\
                           & Vorbis & level 7    &  0.0 & 100 & -0.1 &  -16.75 \\
        \midrule
        \mrow{6}{=}{\LMD}  & FLAC   & level 6    &  0.2 & 100 &  0.2 & -121.33 \\
                           & MP3    & \kbps{320} & -3.1 &  60 & -2.0 &   -2.04 \\
                           & MP3    & \kbps{128} & -3.5 &  62 & -2.4 &   -2.40 \\
                           & AAC    & \kbps{256} & -3.6 &  56 & -1.4 &   -1.03 \\
                           & AAC    & level 5    & -3.8 &  65 & -1.4 &   -1.03 \\
                           & Vorbis & level 7    &  0.0 & 100 & -0.1 &  -17.56 \\
        \midrule
        \mrow{6}{=}{\FP}   & FLAC   & level 6    &  0.2 & 100 &  0.2 & -126.49 \\
                           & MP3    & \kbps{320} & -3.2 &  71 & -0.6 &   -3.04\\
                           & MP3    & \kbps{128} & -3.6 &  71 & -0.2 &   -3.33 \\
                           & AAC    & \kbps{256} & -3.6 &  63 & -0.2 &   -2.35 \\
                           & AAC    & level 5    & -3.5 &  63 & -0.2 &   -2.34 \\
                           & Vorbis & level 7    &  0.0 & 100 & -0.2 &  -17.28 \\
        \midrule
        \mrow{6}{=}{\SEWF} & FLAC   & level 6    &  0.2 & 100 &  0.2 & -135.91 \\
                           & MP3    & \kbps{320} & -3.5 &  29 & -0.2 &   -1.14 \\
                           & MP3    & \kbps{128} & -3.6 &  30 & -3.5 &   -1.43 \\
                           & AAC    & \kbps{256} & -3.7 &  27 & -0.2 &   -0.46 \\
                           & AAC    & level 5    & -3.8 &  27 & -0.2 &   -0.46 \\
                           & Vorbis & level 7    &  0.0 & 100 & -0.1 &  -16.74 \\
        \bottomrule
    \end{tabularx}
    \caption[BPEAQ ODG, 2f, and APEAQ ODG Scores plus BPEAQ's \TNMR for Various Test Tracks]{Scores for basic PEAQ ODG, 2f model, and advanced PEAQ ODG as well as basic PEAQ \TNMR values for all tested audio tracks in \cref{tbl:evaluation:codec+compression+dataset}.}
    \label{tbl:evaluation:dataset}
\end{table}

Our dataset was then tested using multiple PEAQ experiments and overall results for a variety test tracks are shown in \cref{tbl:evaluation:dataset}. As expected, lossless FLAC achieves perfect scores under all metrics. This means FLAC's perceptual quality is near-identical to the original audio's perceptual quality. Both MP3 \kbps{320} and MP3 \kbps{128}, along with AAC \kbps{256} and AAC VBR level 5, score near -4 on BPEAQ, indicating very perceptible impairment (\cf\ \cref{tbl:setup:peaq:odg}). The 2f score of both MP3 codecs as well as both AAC codecs range from 27 (poor) to 71 (fair) indicating that, if these audio codecs were tested in MUSHRA, trained listeners would find these four audio codecs somewhat decent on some tracks and below average on other tracks.
\par

The observed differences in 2f scores may partly reflect variations in the spectral and temporal characteristics of the source material. A pop song like Fleetwood Mac's {\em Dreams} as well as material from the disco genre, \eg Earth, Wind \& Fire's {\em September}, contain dense rhythmic instrumentation including prominent percussion and sustained mid–to-high frequency energy, which increase temporal modulation complexity and may amplify codec-induced modulation distortion. In contrast, a classical performance such as Debussy's {\em La mer} exhibits more structured orchestral textures with comparatively smoother spectral evolution. Rock performances like Pallbearer's {\em Foreigner} contain heavily distorted guitar signals whose broadband harmonic content can raise masking thresholds and partially conceal additional codec artifacts. These genre dependent signal properties may contribute to the better 2f scores observed for the latter tracks.
\par

Using APEAQ, most encoded tracks show good ODG score of 0 and -1. The only exceptions are Debussy's {\em La mer}, scoring at -2 for MP3 \kbps{320} and MP3 \kbps{128}, and Earth, Wind \& Fire's {\em September}, scoring worst at -3.5 for MP3 \kbps{128}, indicating perceptible impairment (\cf\ \cref{tbl:setup:peaq:odg}). This indicates that the compression impairment in MP3 and AAC audio codecs may not be as severe as previously shown by BPEAQ on some tracks. However, audio impairment of Earth, Wind \& Fire's {\em September} encoded with MP3 \kbps{128} is rated in both BPEAQ and APEAQ as quite perceptible. \TNMR values for both MP3 audio-codec settings as well as both AAC audio-codec settings range from -3 to 0 meaning the noise to mask ratio is rather high and most noise from these encoded test tracks will be audible to listeners.
\par

The most interesting case in \cref{tbl:evaluation:dataset} is Vorbis VBR level 7. It is the only lossy audio codec that scores 0 on both BPEAQ ODG and APEAQ ODG as well as 100 on 2f score. This strongly indicates Vorbis' audio impairment is imperceptible to listeners and the audio codec's perceptual quality is very close to lossless FLAC perceptual quality. Vorbis' \TNMR value is around -17, which means its noise level is less audible than the remaining tested audio codecs' noise level.
\par

\begin{table}
    \newcommand{\hl}[1]{\bfseries #1}
    \centering\sffamily\smaller
    \begin{tabularx}{\linewidth}{L{0.5}L{1.7}R{0.9}R{1}R{0.9}R{1}}
        \toprule
        \mrow{2}{=}{Codec} &
        \mrow{2}{=}{Setting} &
        \mrow{2}{=}{BPEAQ ODG} &
        \mrow{2}{=}{2f Score} &
        \mrow{2}{=}{APEAQ ODG} &
        \mrow{2}{=}{BPEAQ \TNMR} \\
         & & & & & \\ 
        \midrule
        FLAC   & level 6        & \hl{0.2} & \hl{100} & \hl{0.2} & \hl{-130.11} \\
        MP3    & CBR \kbps{320} &    -3.6  &      36  &    -0.2  &       -1.34  \\
        MP3    & CBR \kbps{128} &    -3.7  &      36  &    -0.4  &       -1.72  \\
        AAC    & CBR \kbps{256} &    -3.7  &      32  &    -0.4  &       -0.08  \\
        AAC    & VBR level 5    &    -3.8  &      37  &    -0.4  &       -0.08  \\
        Vorbis & VBR level 7    &     0.0  &      99  &    -0.1  &      -16.74  \\
        \bottomrule
    \end{tabularx}
    \caption[Average B.ODG, 2f, A.ODG, plus BPEAQ's \TNMR for Tested Audio Codecs]{Average scores for basic PEAQ ODG, 2f model, and advanced PEAQ ODG as well as PEAQ \TNMR values for all tested audio codecs (\cf\ \cref{tbl:evaluation:odg+avgs}). Best performance in each category has been \hl{marked}.}
    \label{tbl:evaluation:odg+avgs}
\end{table}

\Cref{tbl:evaluation:odg+avgs} presents the average BPEAQ ODG scores, average 2f scores, average APEAQ ODG scores as well as average BPEAQ \TNMR{}s for all tested audio codecs using the quality settings defined in \cref{tbl:evaluation:odg+avgs}. BPEAQ ODG for FLAC is 0 and for Vorbis 0.2 on average, translating to perceptual quality (nearly) the same as the original audio signal. The remaining audio codecs' BPEAQ ODG average close to -4, which means theses audio codecs' impairment is classified as {\em very annoying} (\cf\ \cref{tbl:setup:peaq:odg}). These partial results are supported as well by the average 2f-model scores, which indicate that FLAC and Vorbis would score 100 on the MUSHRA listening test while the remaining codecs perform at much lower values in the 30s. The average \TNMR also shows that noise in all lossy audio codecs is very noticeable except for Vorbis. The interesting case here is the APEAQ ODG score, which shows much better values for all audio codecs at near 0. This indicates that APEAQ regards all tested audio codecs impairment as imperceptible, which contradicts with all the BPEAQ ODG scores, 2f-model scores, and \TNMR values.
\par

\subsection{Machine-Learning Based Codec}
\label{sec:evaluation:rvqgan}

In recent years, a trend emerged to apply machine-learning techniques into remastering and compressing audio. One proposed audio compression model employs a residual vector quantized generative adversarial network (RVQGAN) \autocite{Kumar:HFACIRVQGAN:2023}. RVQGAN works by encoding the audio signal into a lower-dimensional presentation followed by conversion into discrete values. We employed RVQGAN's prototype implementation from its \textcite{DACRVQGANGithubRepo} and encoded the Black Sabbath's {\em Iron Man} with it.
\par

\begin{table}
    \newcommand{\hl}[1]{\bfseries #1}
    \centering\sffamily\smaller
    \begin{tabularx}{\linewidth}{L{1}L{.7}L{1.15}R{1.25}R{.95}R{.95}}
        \toprule
         &
        \mrow{2}{=}{Codec} &
        \mrow{2}{=}{Setting} &
        \mrow{2}{=}{Compression Ratio} &
        \mrow{2}{=}{Encoding Speed} &
        \mrow{2}{=}{Decoding Speed} \\
         & & & & & \\
        \midrule
        Lossless           & FLAC   & level 6    &     38.74  & \hl{31.92} &      92.47  \\
        \midrule
        \mrow{5}{=}{Lossy} & MP3    & \kbps{320} &     83.99  &      4.90  &     112.30  \\
                           & MP3    & \kbps{128} &     92.44  &      5.70  & \hl{148.93} \\
                           & AAC    & \kbps{256} &     86.73  &      7.12  &      33.07  \\
                           & AAC    & level 5    &     87.42  &      6.03  &      31.97  \\
                           & Vorbis & level 7    &     88.68  &      8.69  &      46.39  \\
        \midrule
        AI                 & RVQGAN &            & \hl{98.15} & 1.45       &       1.03  \\
        \bottomrule
    \end{tabularx}
    \caption[Average Audio Codec Compression Ratios plus RVQGAN]{Average audio-codec compression as percent ratio and encoding as well as decoding speeds in samples per microsecond. Values for lossless and lossy encoding are the same as in \cref{tbl:evaluation:codec+compression+speed} with an additional result row for RVQGAN. Best performance in each category has been \hl{marked}.}
    \label{tbl:evaluation:codec+compression+speed+rvqgan}
\end{table}

\Cref{tbl:evaluation:codec+compression+speed+rvqgan} presents RVQGAN's average compression ratio, average encoding speed, and average decoding speed in comparison with traditional audio codecs previously evaluated. In its current state, RVQGAN does improve on the compression ratio by reducing storage size to almost 1\% of the original size. However, encoding and decoding speed are nearly one to two magnitudes slower than any of the traditional audio codecs.
\par

\begin{figure}
    \centering
    \includegraphics[width=\linewidth]{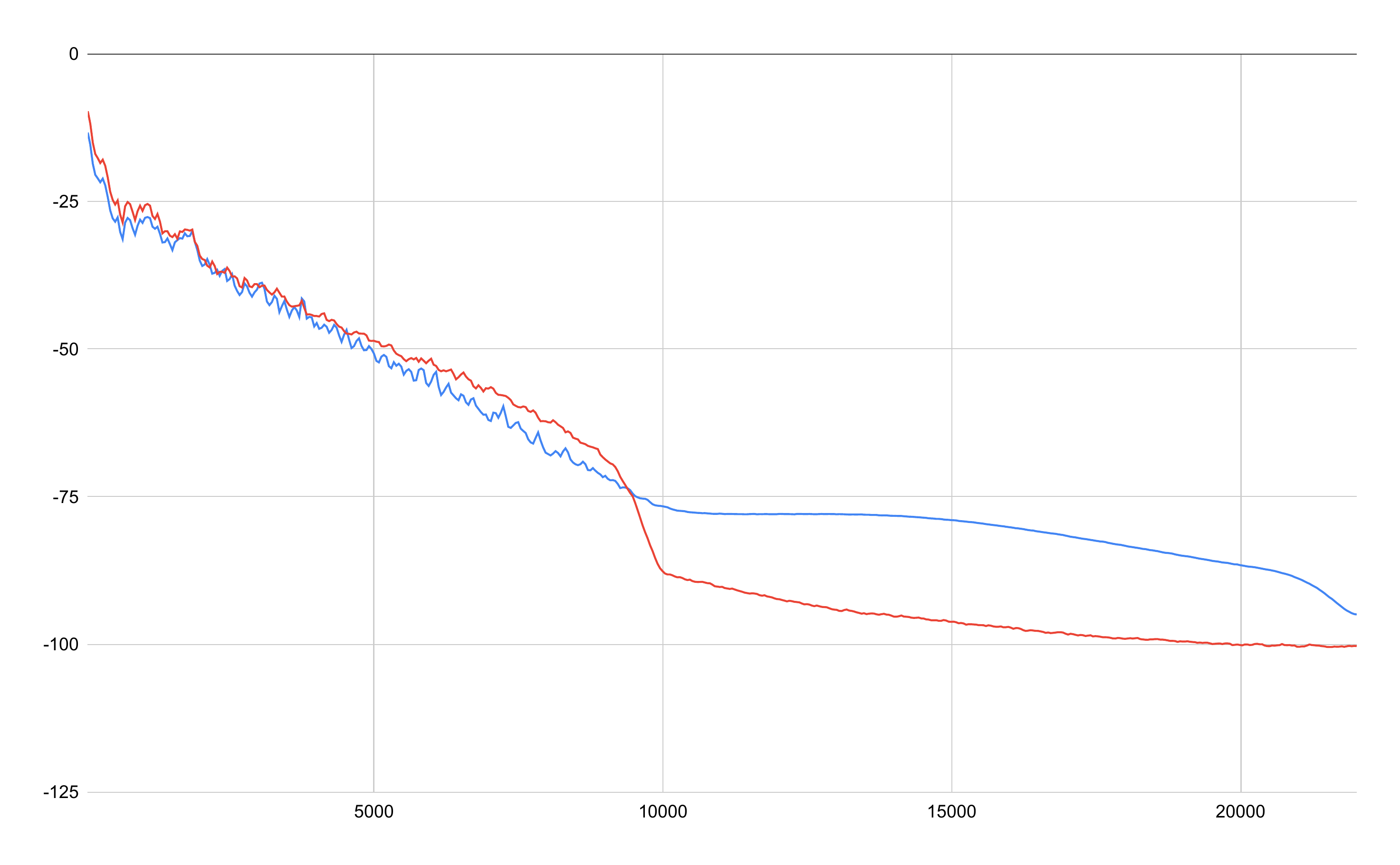}%
    \caption[RVQGAN Spectrum]{RVQGAN's spectrum (red) together with the uncompressed signals spectrum (blue) (exported from Audacity). X-axis shows the frequency range of the sample track in \HZ{}, y-axis shows loudness range from \DB{0} to \DB{-125}. Compared to the uncompressed audio signal, the RVQGAN encoded audio signal shows a slight increase in loudness before dropping off at around \KHZ{10}, gradually reducing in volume towards higher frequencies.}
    \label{fig:evaluation:spectrum+rvqgan}
\end{figure}

\begin{figure}
    \centering
    \subfloat[][Uncompressed]{%
        \includegraphics[trim=0 0 0 50,clip,width=0.492\linewidth]%
        {figures/Iron_Man_og.png}%
        \label{fig:evaluation:iron+man:og}
    }\subfloat[][RVQGAN]{%
        \includegraphics[trim=0 0 0 50,clip,width=0.492\linewidth]%
        {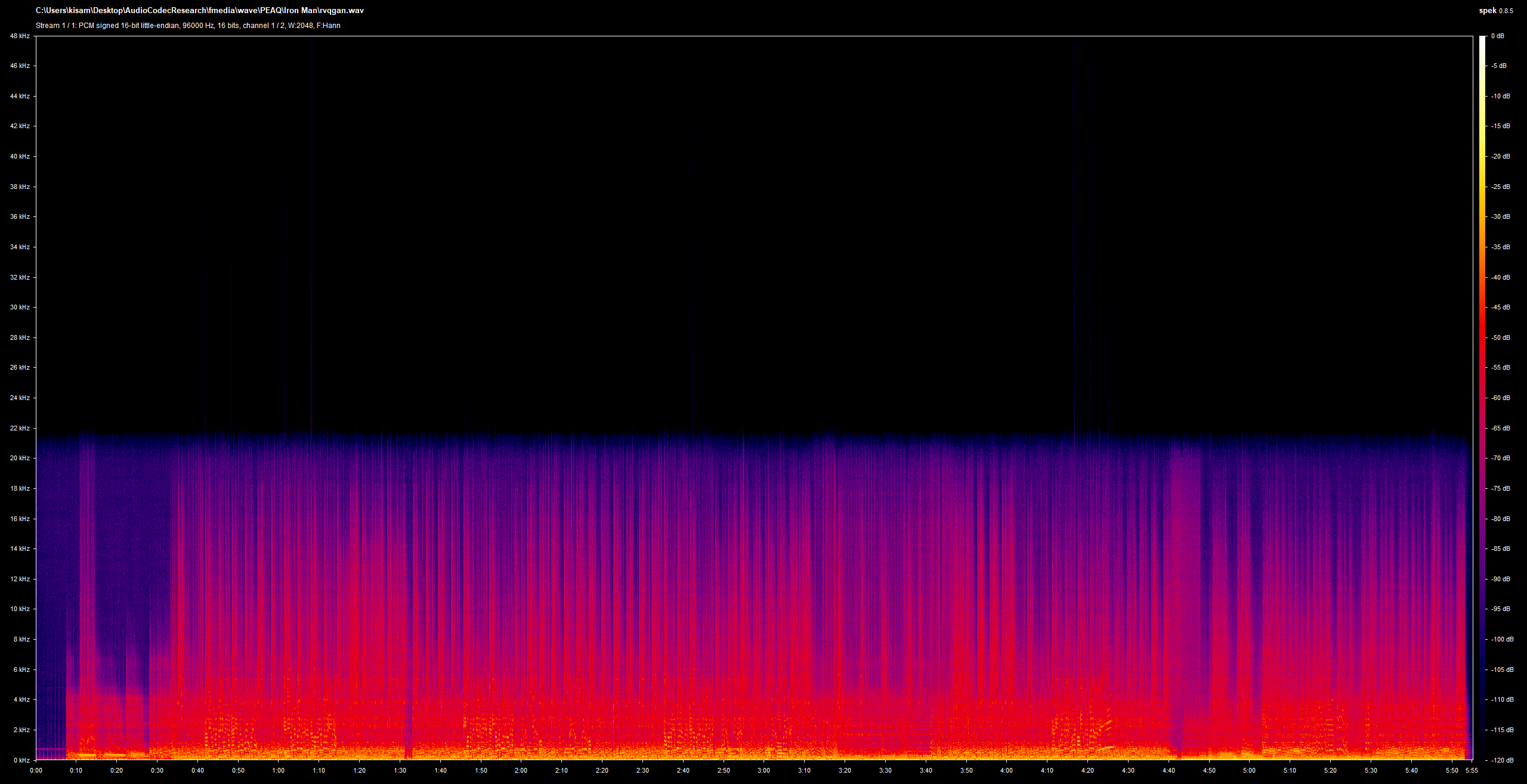}%
        \label{fig:evaluation:iron+man:rvqgan}
    }
    \caption[Audible Loudness Comparison (Uncompressed/RVQGAN)]{Audible loudness comparison for the track {\em Iron Man} by Black Sabbath compressed using \protect\subref{fig:evaluation:iron+man:rvqgan} RVQGAN. Compared to the uncompressed audio signal in \protect\subref{fig:evaluation:iron+man:og}, the RVQGAN encoded audio signal exhibits a cut-off in loudness at around \KHZ{21}. X-axis shows duration of the tested audio signal, y-axis shows the frequency range of the tested audio signal. Color spectrum indicates loudness in \DB{}.}
    \label{fig:evaluation:iron+man+rvqgan+spek}
\end{figure}

\begin{figure}
    \centering
    \subfloat[][Uncompressed]{%
        \includegraphics[width=0.492\linewidth]{figures/SF-og.png}%
        \label{fig:evaluation:iron+man:og+sf}
    }\subfloat[][RVQGAN]{%
        \includegraphics[width=0.492\linewidth]{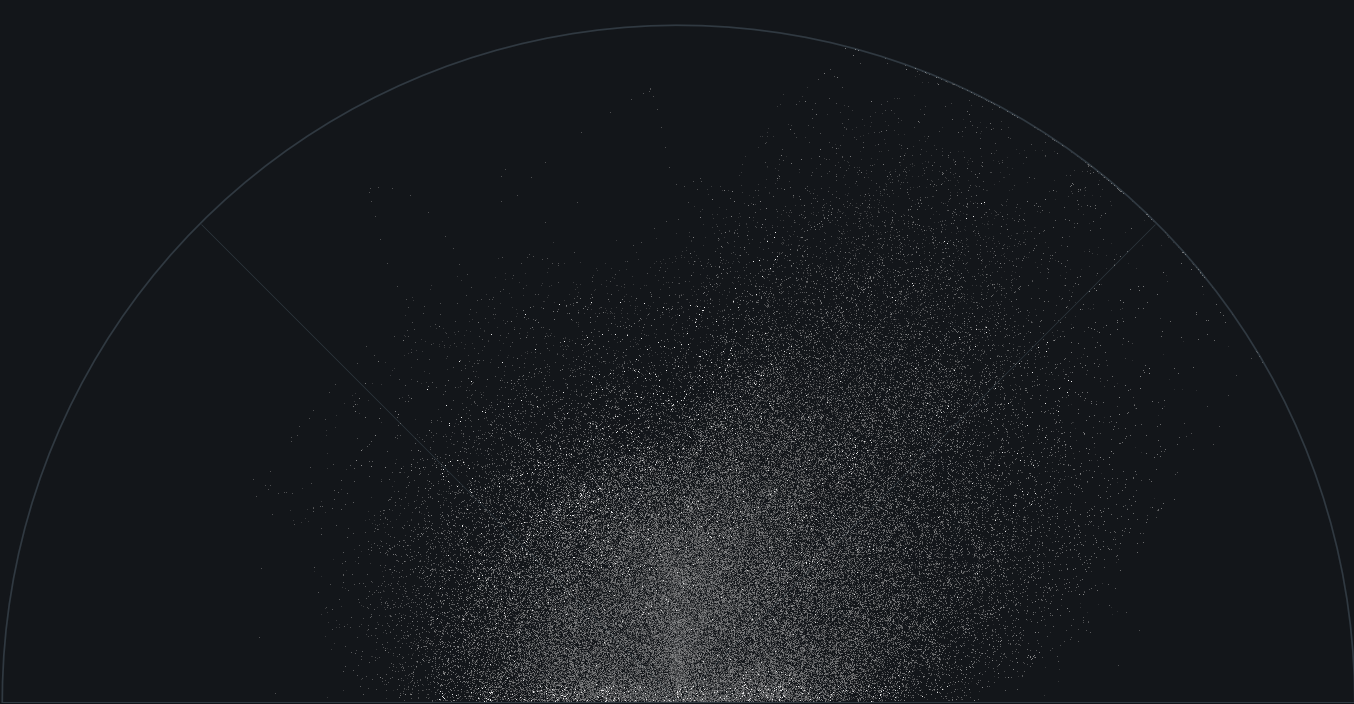}%
        \label{fig:evaluation:iron+man:rvqgan+sf}
    }
    \caption[Sound-field visualizations (RVQGAN)]{Sound-field visualizations at \MIN{3:00\,--\,3:30} for the track {\em Iron Man} by Black Sabbath. \protect\subref{fig:evaluation:iron+man:og+sf} uncompressed recording and \protect\subref{fig:evaluation:iron+man:rvqgan+sf} compressed with RVQGAN. It is apparent that the pattern in the RVQGAN encoded audio signal output is more {\em smeared} and lacks the accuracy of the guitar's position when compared to the uncompressed audio's sound field.}
    \label{fig:evaluation:iron+man+rvqgan+sf}
\end{figure}

\Crefrange{fig:evaluation:spectrum+rvqgan}{fig:evaluation:iron+man+rvqgan+sf} illustrate the spectrum, spectrogram, and sound-field visualization of an RVQGAN encoded audio signal compared to the original uncompressed audio signal. The spectrogram in \cref{fig:evaluation:iron+man+rvqgan+spek} shows a cutoff in loudness at around \KHZ{21}. The spectrum in \cref{fig:evaluation:spectrum+rvqgan} shows the RVQGAN tested track exhibits first a slight loudness increase in the lower frequency range, which gradually decreases form around \KHZ{10} to \DB{-100} at \KHZ{20} and beyond. \cref{fig:evaluation:iron+man+rvqgan+sf} shows the encoded audio-signal output pattern more smeared and less accurate compared to the uncompressed track's original pattern. Overall, these visualizations indicate that RVQGAN encoded audio signals may share similar perceptual quality problems with lossy encoded audio.
\par

\begin{table}
    \newcommand{\hl}[1]{\bfseries #1}
    \centering\sffamily\smaller
    \begin{tabularx}{\linewidth}{L{0.5}L{1.7}R{0.9}R{1}R{0.9}R{1}}
        \toprule
        \mrow{2}{=}{Codec} &
        \mrow{2}{=}{Setting} &
        \mrow{2}{=}{BPEAQ ODG} &
        \mrow{2}{=}{2f Score} &
        \mrow{2}{=}{APEAQ ODG} &
        \mrow{2}{=}{BPEAQ \TNMR} \\
         & & & & & \\ 
        \midrule
        FLAC   & level 6        & \hl{0.2} & \hl{100} & \hl{0.2} & \hl{-130.11} \\
        MP3    & CBR \kbps{320} &    -3.6  &      36  &    -0.2  &       -1.34  \\
        MP3    & CBR \kbps{128} &    -3.7  &      36  &    -0.4  &       -1.72  \\
        AAC    & CBR \kbps{256} &    -3.7  &      32  &    -0.4  &       -0.08  \\
        AAC    & VBR level 5    &    -3.8  &      37  &    -0.4  &       -0.08  \\
        Vorbis & VBR level 7    &     0.0  &      99  &    -0.1  &      -16.74  \\
        \midrule
        RVQGAN &                &    -2.7  &      37 &     -3.7  &       -0.69 \\
        \bottomrule
    \end{tabularx}
    \caption[Average B.ODG, 2f, A.ODG, plus BPEAQ's \TNMR plus RVQGAN for Tested Audio Codecs]{Average scores for basic PEAQ ODG, 2f model, and advanced PEAQ ODG as well as PEAQ \TNMR values for all tested audio codecs (\cf\ \cref{tbl:evaluation:odg+avgs}) plus average scores for RVQGAN. Best performance in each category has been \hl{marked}.}
    \label{tbl:evaluation:odg+rvqgan}
\end{table}

Despite achieving an impressive compression ratio of around 98\% of the uncompressed file size, perceptual quality of RVQGAN encoded audio is questionable. Currently, RVQGAN exhibits worse perceptual quality than all traditional audio codecs. RVQGAN's average evaluation scores in \cref{tbl:evaluation:odg+rvqgan} supports this evaluation.
\par

\section{Discussion}
\label{sec:discussion}

\subsection{Audio-Codec Performance}
\label{sec:discussion:audio+codec+performance}

In our audio-codec performance tests (\cf\ \cref{tbl:evaluation:codec+compression+speed}), FLAC exhibited the lowest compression ratio of all tested audio codecs meaning that lossless audio files will take up more storage space than files encoded using a lossy compression technique. On the other hand, FLAC achieved the highest encoding speed. These two average values for FLAC are expected because lossless compression techniques preserve most of the information in an audio signal while only compressing redundancies and repeated patterns.
\par

All lossy audio codecs have similar compression ratios with MP3 \kbps{128} achieving the best results. Both MP3 quality settings, \ie\ \kbps{128} and \kbps{320}, exhibit lower encoding speeds than other lossy audio codecs. On the other hand, these codecs show significantly higher decoding speeds than other lossy audio codecs with MP3 \kbps{128} exhibiting the fastest decoding speed. It can be argued that MP3 is a strong contender for the best lossy audio codec due to its high decoding speeds, making it the ideal choice for audio playback on low-power devices. This is because MP3 has lower computational complexity compared to modern codecs like AAC or Vorbis, which is due to its older more simple perceptual encoding model \autocite{Brandenburg:MP3AACE:1999}. However, MP3's encoding speed is constrained by its {\em  two-nested-iteration-loop} design \autocite{Brandenburg:MP3AACE:1999}, which makes MP3's encoding speed fall behind when comparing to AAC and Vorbis newer and more sophisticated encoding techniques.
\par

Among lossy audio encoders, Vorbis exhibits above average compression ratios, the fastest encoding speeds, and decent decoding speeds. In our audio-codec performance evaluation, Vorbis is the runner up for lossy audio codecs, providing good compression ratios and fastest lossy encoding speeds.
\par

\subsection{Perceptual Quality}
\label{sec:discussion:perceptual+quality}

Our perceptual quality evaluation shows that, while lossy compression algorithms still try to preserve perceptual audio quality, they sacrifice high-frequency aural details to reduce storage size, thus reducing the quality of the audio signal. Because audio encoding using psycho-acoustic principles determines regions that are mostly inaudible to the average listener, audio data for these regions are simply removed. The spectra for all lossy audio codecs (\cf\ \cref{fig:evaluation:spectrum}) clearly show how each lossy audio codec handles loudness reduction of high-frequency regions.
\par

Loudness reduction and removal of high frequencies reduces the harmonic richness, the details and clarity, and the spatial depth \autocite{Truax:THFAE:2020} that allow digital music to exhibit high-fidelity audio qualities. \Textcite{Vickers:TLWBSR:2010}, mainly reporting on the {\em loudness war} in music production, also shows that the average human listener is able to perceive aural quality very different depending on the loudness alteration in a music recording. Additionally, loudness reduction and high-frequency removals negatively affect the stereo image of the encoded audio signal reducing aural accuracy (\cf\ \cref{fig:setup:visual:sound+field}).
\par

We employ multiple complementary PEAQ predictors, including basic PEAQ, advanced PEAQ, basic PEAQ's \TNMR values, and the 2f model. Using a combination of not just ODG scores but also MOV scores allows to capture distinct psychoacoustic dimensions. Convergence across these metrics (\cf\ \cref{tbl:evaluation:odg+avgs}) strengthens confidence that observed audio-codec artifacts align with listener perceptible phenomena, providing a reproducible computational approximation rather than a substitute for subjective evaluation.
\par

The ODG scores and the MOV scores collected from all tested audio compression codecs (\cf\ \cref{tbl:evaluation:odg+avgs}) support these findings. On average, all traditional audio codecs, \ie FLAC, MP3, AAC, and Vorbis, score near 0 on advanced PEAQ ODG. This means that advanced PEAQ classifies all of these traditional audio codecs' impairment level as imperceptible (\cf\ \cref{tbl:setup:peaq:odg}). However, the noise level of lossy audio codecs is higher than the noise level of lossless audio codecs. In particular, noise levels for MP3 and AAC are right at the threshold of their masking levels, meaning their noise levels are potentially perceptible. Remarkably, Vorbis is the only lossy audio-compression codec where the noise level is masked reasonably well, making it the winner in this category. The 2f-model scores support this as well with Vorbis' achieving an average score of 99 while other lossy encoders' average 2f-model scores are in the 30s.
\par

The notable discrepancy between basic PEAQ ODG scores and advanced PEAQ ODG scores observed for encoded audio with MP3 and AAC (\cf\ \cref{tbl:evaluation:odg+avgs}) warrants further discussion. While basic PEAQ yields an average ODG score of approximately -3.7 for both codecs, suggesting very annoying impairment, advanced PEAQ in contrast produces scores near 0, suggesting imperceptible impairment. This gap can be attributed to the difference in perceptual modeling depth between the two models. Basic PEAQ's MOVs, which rely primarily on noise-to-mask ratio and loudness-based metrics \autocite{Thiede:PEAQITUSOMPAQ:2000}, can be rather rigid when evaluating audio artifacts above basic PEAQ's masking threshold, even if not be perceptible for a human listener. Advanced PEAQ, by contrast, incorporates MOVs capturing modulation masking, harmonic coherence, and asymmetric distortion weighting \autocite{Thiede:PEAQITUSOMPAQ:2000}, which together better account for the structured, partially-masked nature of audio encoding artifacts. 
\par

The above findings suggests that basic PEAQ's ODG score should be interpreted with caution when evaluating modern perceptual codecs. The magnitude of the gap between basic ODG scores to advanced ODG scores in itself may serve as an indicator of how well a codec's artifact profile aligns with a given perceptual model's assumptions. Our finding of a gap between basic PEAQ ODG scores and advanced PEAQ ODG scores also agrees with work by \textcite{Delgado:CWSUP:2020}, which motivated conducting additional evaluation with the 2f model and basic PEAQ's \TNMR MOV. Together these four metrics allow a comprehensive and objective comparison of an audio codec's perceptual quality characteristics.
\par

While most of our assessment was conducted via a system of audio codec performance measurements, visualizations, and various PEAQ-related tests, we have to address the importance of subjective listening tests. PEAQ variants were built under \textcite{ITU:BS.1387-2:2023} to predict how an average human listener perceives audio quality \autocite{ITU:BS.1387-2:2023}. Trained listeners, under \textcite{ITU:BS.1534-3:2015} conducting MUSHRA tests, are obligated to meet certain (minimum) requirements to qualify as {\em assessors} \autocite{ITU:BS.1534-3:2015}. Thus, trained listeners may be able to provide additional feedback about how humans perceive aural quality, which would be useful to validate the results of the objective evaluation models and, hence, validate and improve our study and approach.
\par

\subsection{AI-based Audio Codecs}
\label{sec:discussion:ai-based+audio+codecs}

Our tests indicate RVQGAN \autocite{Kumar:HFACIRVQGAN:2023} indeed exhibits the highest compression ratio compared with traditional audio codecs (\cf\ \cref{tbl:evaluation:codec+compression+speed+rvqgan}). Storage size of an RVQGAN encoded audio signal is only around 1\% to 2\% of the size of the uncompressed signal, which is an impressive result. However, RVQGAN's encoding and decoding speeds are very low compared to traditional audio codecs, performing one to two magnitudes slower. Additionally, in its current version RVQGAN's output is a unique file format, {\em descript-audio-codec} (\CMD{.dac}) \autocite{DACRVQGANGithubRepo}, which is not playback-supported by any audio player application at the time of our evaluation tests.
\par

While the RVQGAN codec achieved the highest average compression ratio in our audio-codec performance tests, our testing shows that RVQGAN's perceptual quality is at the {\em worse end} of traditional audio codecs' perceptual quality. RVQGAN's average advanced ODG score is -3.7, which is close to {\em very annoying} (\cf\ \cref{tbl:setup:peaq:odg}), and RVQGAN's average \TNMR value indicates that its noise level is quite perceptible (\cf\ \cref{tbl:evaluation:odg+rvqgan}). AI-based approaches to {\em developing} or {\em discovering} audio codecs shows promise in advancing efficiency but will require further refinements to provide acceptable hi-fi listening experiences to end users.
\par

\section{Conclusions}
\label{sec:conclusions}

We evaluated several of the most common audio-compression codecs providing insights in how audio-compression techniques affect sonic perceptual quality of digital music. As expected, our tests indicate that lossy audio-compression codecs do provide better compression performance than lossless audio-compression codecs. This stems mainly from the fact that lossy audio-compression algorithms sacrifice aural perceptual quality for higher compression speeds and lower final storage size. However, the Vorbis audio codec emerged as the exception to this rule. It exhibits acceptable compression efficiency when compared to its competitors while at the same time achieving perceptual quality nearly imperceptibly distinguishable compared to the respective uncompressed digital-audio signal. Finally, if storage size is not a concern, lossless audio-compression codecs such as FLAC will be provider better aural perceptual quality in reproducing the original audio signal.
\par

We also evaluated RVQGAN, a machine-learning based audio compression scheme. While RVQGAN's codec performance in encoding speed and final storage size was impressive, its aural perceptual quality is severely lacking. We are sure to see more advancement in this field, especially improving the perceptual quality of future AI-driven audio-compression codec developments.
\par

Our evaluation strengthens our argument that, when choosing a digital-audio compression codec, not only audio-compression performance should be a focus. Users should also consider the perceptual quality of the encoded audio signal to ensure the best possible listening experience. We compliment Vorbis as an excellent lossy audio-compression codec, which is also able to maintain high-fidelity perceptual quality. The demonstrated consistency between objective evaluation metrics, \ie vasic PEAQ, advanced PEAQ, and the 2f model, and established MUSHRA behavior provides further evidence that objective perceptual modeling can serve as a robust foundation for audio-codec evaluation without reliance on extensive listening panels.
\par

\subsection{Future Work}
\label{sec:conclusions:future+work}

Expanding our database of audio-codec performance measurements can determine the consistency in compression ratio, encoding speed, and decoding speed of all audio-compression codecs. We welcome additional performance-measurement submissions to our database {\smaller\url{https://github.com/samcate/fmedia-Audio-Codec-Comparison-Database}} by the community.
\par

With the recent trend of increased use of high-fidelity music streaming services like \textcite{Spotify}, perceptual quality evaluation should be extended to streaming audio as well. Streaming services utilize a wide range of audio-compression schemes, which can provide interesting insights, especially when employing objective scoring.
\par

Future research should also focus on enhancing the perceptual quality of AI-driven compression methods. By integrating psycho-acoustic principles into a compression neural network it may be possible to achieve more accurate perceptual quality in addition to providing low-storage size high-fidelity digital music to users.
\par



\printbibliography[title={References}]

\end{document}